\documentclass[a4paper]{article}

\usepackage{epsfig}
\usepackage{times, courier}

\usepackage{amsfonts}

\usepackage{paralist} 

\long\def\comment#1{}

\newlength{\figtblfootnotemargin}
\newlength{\figtblfootnotewidth}

\newcommand{\SquareOfSize}[2]{
 \fbox{\hsize #1cm \hbox to #1cm{\vbox{#2}}}
}

\newsavebox{\wholeWidthLine}
\sbox{\wholeWidthLine} {\rule[0.1in]{\textwidth}{.01in}}

\newcommand{\bq}{\begin{quote}}
\newcommand{\eq}{\end{quote}}
\newcommand{\be}{\begin{enumerate}}
\newcommand{\ee}{\end{enumerate}}
\newcommand{\bi}{\begin{itemize}}
\newcommand{\ei}{\end{itemize}}
\newcommand{\bie}{\begin{itemize}\begin{enumerate}}
\newcommand{\eie}{\end{enumerate}\end{itemize}}
\newcommand{\ba}{\begin{array}}
\newcommand{\ea}{\end{array}}
\newcommand{\btbl}{\begin{tabular}}
\newcommand{\etbl}{\end{tabular}}
\newcommand{\bequ}{\begin{displaymath}}
\newcommand{\eequ}{\end{displaymath}}
\newcommand{\bequa}{\begin{eqnarray*}}
\newcommand{\eequa}{\end{eqnarray*}}
\newcommand{\bc}{\begin{center}}
\newcommand{\ec}{\end{center}}
\newcommand{\btab}{\begin{tabbing}}
\newcommand{\etab}{\end{tabbing}}

\newcommand{\godown}{ \vspace*{0.3cm}}

\def\mpr#1{\ifmmode #1 \else #1 \fi}

\newcommand{\DPATTRNAME}{{\large{{\bf SC}$_{attr}$}}}
\newcommand{\DPISANAME}{{\large{{\bf SC}$_{isA}$}}}
\newcommand{\DPINNAME}{{\large{{\bf SC}$_{in}$}}}

\newtheorem{myproposition}{{\bf Prop.}}
\newtheorem{example}{Example}[section]

\newcommand{\B}{\mathbb{B}}

\newcommand{\OU}{{\cal O}}

\newcommand{\ic}{{\tt IC}}
\newcommand{\poi}{{\tt POI}}
\newcommand{\cpoi}{{\tt CPOI}}
\newcommand{\cpoiu}{{\tt CPOI$_U$}}
\newcommand{\cb}{{\tt CB}}
\newcommand{\tcb}{{\tt TSB}}
\newcommand{\cbd}{{\tt CBD}}
\newcommand{\DatA}{{\tt Dat1}}
\newcommand{\DatB}{{\tt Dat2}}
\newcommand{\DatC}{{\tt Dat3}}

\begin{document}

\title{CPOI: A Compact Method to Archive \\ Versioned RDF Triple-Sets}

\comment{
    \numberofauthors{1}
    \author{
    \alignauthor
    Maria Psaraki and Yannis Tzitzikas\\
        \affaddr{Institute of Computer Science, FORTH-ICS, and}\\
        \affaddr{Computer Science Department, University of Crete}\\
        \affaddr{Heraklion, Crete, GREECE}\\
        \email{\{psaraki$|$tzitzik\}@ics.forth.gr}
    }
===}

\comment{
    \author{Maria Psaraki and Yannis Tzitzikas}
        \institute{Computer Science Department, University of Crete, GREECE, and\\
                Institute of Computer Science, FORTH-ICS, GREECE\\
                email: {\tt \{psaraki, tzitzik\}@ics.forth.gr}
        }
}

\comment{
\author[M. Psaraki et al]{Maria Psaraki and
                            Yannis Tzitzikas \\
Institute of Computer Science, FORTH-ICS, Greece, and \\
Computer Science Department, University of Crete, Greece}
}

\author{Maria Psaraki and Yannis Tzitzikas \\
Institute of Computer Science, FORTH-ICS, Greece, and \\
Computer Science Department, University of Crete, Greece}

\maketitle
\begin{abstract}
Large amounts of RDF/S data are produced and published lately,
and several modern applications  require the provision of versioning and archiving services
over such datasets.
In this paper we propose a novel storage index for archiving versions of such datasets,
called
\cpoi\ (compact partial order index),
that exploits the fact that an RDF Knowledge Base (KB),
is a graph (or equivalently a set of triples), 
and thus it has not a unique serialization
(as it happens with text).
If we want to keep stored several versions
we actually want to store multiple sets of triples.
\cpoi\ is a data structure for storing such sets
aiming at reducing 
the storage space
since this is important not only for
reducing storage costs,
but also for
reducing the various communication costs
and
enabling hosting in main memory (and thus processing efficiently) large quantities of data.
\cpoi\ is based on a {\em partial order structure}
over sets of triple identifiers,
where
the triple identifiers
are represented
in a {\em gapped form}
using {\em variable length} encoding schemes.
For this index we evaluate analytically and experimentally various
identifier assignment techniques
and their space savings.
The results show significant  storage savings,
specifically,
the storage space of the compressed sets
in large and realistic synthetic datasets
is about  the 8\% of the size of the uncompressed sets.
\end{abstract}



\section{Introduction}

The rising tide of data is a main characteristic of our age:
``{\em We create as much information
in two days now as we did from the dawn of man through 2003}".\footnote{E. Schmidt (CEO of Google), 2010, http://techcrunch.com/2010/08/04/schmidt-data/}
%
A large proportion of these data are scientific.
For instance,
and according to \cite{eu:data},
the first ``reading" of the human genome
created digital records on more than 250 billion DNA bases in less than 10 years,
while
the evolution of European Space Agency's Earth Observation data archives
passed three petabytes in 2007 (the projection for 2020 is a seven-fold rise).
For scientific data the provision of {\em versioning} services is important
for various purposes (archiving, preservation, provenance).
For instance, failure to keep the previous states of scientific data
(over which other experiments were based)
jeopardizes scientific evidence and our ability to
verify  findings \cite{buneman2004asd}.

Lately, an increasing amount of data (including scientific data)
is published on the Web in RDF/S according
to the Linked Open Data (LOD) principles \cite{berners2009linked}
and various  applications,
e.g. in life sciences \cite{DBLP:conf/dils/HartungKR08,DBLP:journals/jbi/GellerPHC09},
require the provision of {\em versioning} services
over RDF datasets (either schema-free or schema-based).
It follows that versioning services over large amounts of structured (scientific)
data is a requirement from which we cannot escape.

Two key
performance aspects of a version management system is the storage
space and the time for creating (resp. retrieving) a new (resp.
existing) version.
Most of the related works in the SW (Semantic Web) focus on high
level functions
and they mainly overlook the storage space perspective which is fundamental.
It should be stressed
that the space perspective is important
not only for
(a) reducing storage costs,
but also for
(b) reducing the various communication costs
(e.g. loading times from disk or across networks),
and
(c)
enabling hosting in main memory (and thus processing efficiently) large quantities of data.

\begin{figure}
\centering
    \includegraphics[width=80mm]{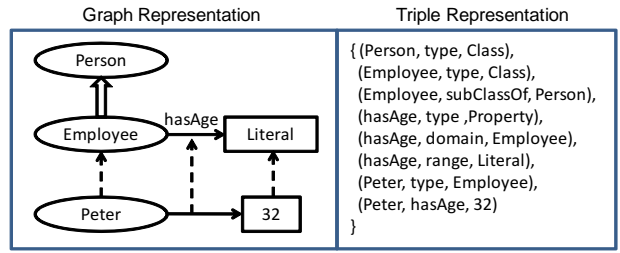}
\vspace*{-0.2cm}
\hspace*{0.9cm}
\caption{An RDF KB as a graph and the corresponding set of triples}
\label{fig:RDF_Graph}
\vspace*{-0.3cm}
\end{figure}

\godown

An RDF/S KB (Knowledge Base) can be viewed as a graph
        or as a set of triples,
        as illustrated in the example of Figure \ref{fig:RDF_Graph}.
Regarding storage, we
can identify the following main approaches of a versions management system. \\
(a) {\em Independent Copies (IC)}:
    Every  version is stored independently and
    this policy is adopted in \cite{SemVersion2005,Noy2004b,klein2002ova}. \\
(b) {\em  Change Based (CB)}:
    Only deltas between subsequent versions are stored,
    a policy that  is adopted in various tools for versioning software
    \cite{Tich85,berliner90cvs},
    and it has been proposed also for Semantic Web data
    \cite{Zeginis2007,AIA2007}.\\
(c) {\em  Timestamp Based (\tcb)}:
    Each triple is enriched with time-stamps
    indicating the versions the triple belongs to
    (\cite{AIA2007,DBLP:journals/pvldb/NeumannW10}).
    Proposals like \cite{DBLP:journals/tkde/GutierrezHV07}
    fall into the same category too.

\begin{figure*}
\centering
    \hspace*{-3mm}
    \includegraphics[width=125mm]{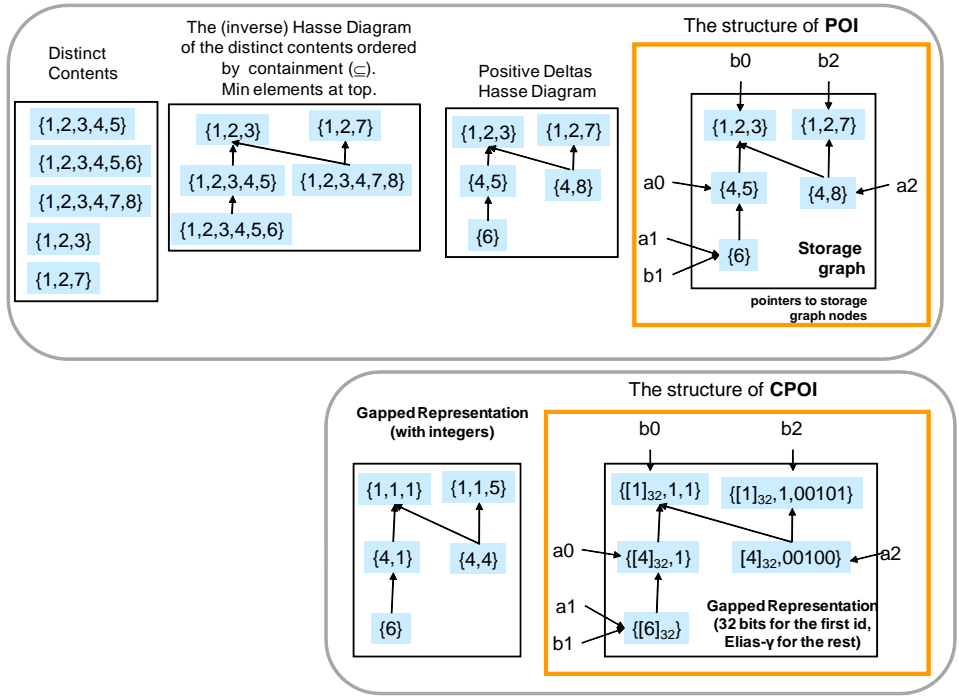}
\vspace*{-0.3cm}
\hspace*{0.9cm}
\caption{The structure of \poi\ and the introduced compact representation \cpoi}
\label{fig:poi_graph}
\vspace*{-0.3cm}
\end{figure*}

In \cite{DBLP:conf/esws/TzitzikasTA08} a storage structure, called
\poi\  (Partial Order Index) has been proposed. \poi\ exploits the
fact that RDF KBs have not a unique serialization (as it happens
with texts) and it offers notable space savings in comparison to the
change-based approach,  as well as efficiency in various cross
version operations. \poi\ views an RDF KB as a set of triples and
exploits the expected overlap between versions' contents in order to
reduce the storage space. We shall describe \poi\ using an example.
\begin{example}\label{ex1}
Consider six (6) KBs, denoted by a0, a1, a2, b0, b1
and b2. Each KB consists of a set of triples and each triple is
assigned a unique numeric identifier.
Specifically suppose that: \\
\btbl{ll}
a0=\{1, 2, 3, 4, 5\}                    & b0=\{1, 2, 3\} \\
a1=\{1, 2, 3, 4, 5, 6\} \ \ \ \ \ \     & b1=\{1, 2, 3, 4, 5, 6\} \\
a2=\{1, 2, 3, 4, 7, 8\}                 & b2=\{1, 2, 7\}
\etbl \ \\
Furthermore suppose that there are two evolution tracks
a0 $\rightarrow$ a1 $\rightarrow$ a2
(meaning that a1 is the next version of a0, and that a2 is the next version of a1),
and
b0 $\rightarrow$ b1 $\rightarrow$ b2.
The upper part of Figure \ref{fig:poi_graph} depicts the structure of \poi.
The first diagram shows
the set of distinct triple sets,
the second diagram
shows the proper subset ($\subset$) relationships that hold between these sets,
while in the third diagram
each node does not contain triples ``inherited" from parent nodes.
This is the {\em storage graph} of \poi.
The fourth diagram shows the entire structure for \poi,
where each version id points to the corresponding
node of the storage graph.
For instance, version a1 points to a node storing only
the triple identifier 6. The full contents of a1 are obtained by
taking the union of the triples stored at that node
and its parent nodes,
i.e. $\{6\} \cup \{4,5\} \cup \{1,2,3\}= \{1,2,3,4,5,6\}$.
Finally,
there is a table $TStrings$ which
maps each triple id to the corresponding triple string,
e.g. $2 \rightarrow ``(Peter, type, Employee)"$. \\
$\diamond$
%
\end{example}

In brief, \poi\ stores explicitly only the triple ids of the versions
with the minimal (with respect to set containment) contents. All the
rest versions are stored in a positively incremental way
which is history-independent.
Past experiments
over synthetic datasets have shown
\cite{DBLP:conf/esws/TzitzikasTA08}
that \poi\ occupies
less space than the change-based approach.
Regarding version retrieval time, the
cost of retrieving the contents of a version in \poi\ is independent
of any kind of history (in comparison with the change-based
approach), but it depends on the contents of the particular version,
specifically on the depth of the corresponding node in the \poi\
graph.

In this paper  we investigate whether techniques which have been used with success
in the area of IR (Information Retrieval) systems and WSE (Web Search Engines)
can be exploited for RDF data.
In IR, the adoption of {\em variable length encoding schemes} in the posting lists of an  inverted index
offers significant space savings.
Specifically, if the documents that share a lot of common words get  close identifiers,
and we adopt a {\em gapped representation}
(explained in the continuation of the running example)
for their identifiers using an encoding scheme
that represents small integers with a small number of bits,
then the postings lists of these common words will have small bit representation.
Experiments in the IR domain
(specifically in the TREC web data \cite{DBLP:conf/sigir/ScholerWYZ02})
have shown that
the \emph{compression ratio},
defined as the size of the compressed
file as a percentage of the uncompressed file
(i.e. $Compression\ ratio = \frac{Compressed\ size}{Uncompressed\ size}100\%$),
is around 30\%-45\%.
In our case,  since each triple has a unique identifier, and
hence each node of \poi\ stores a set of identifiers, we investigate
whether special identifier assignments and integer encodings, such
as
Elias-$\gamma$ \cite{Elias},
can reduce (and under what conditions) the occupied space.
Although \poi\ by construction  offers significant storage gains, the motivation for
investigating this approach
is not only for reducing the space requirements of \poi\
in the expected application scenarios,
but also for tackling
the cases where there are several overlapping versions
which are not  related by inclusion.
In such cases, \poi\ behaves
like the IC approach.
It would be desirable to have an index structure
that behaves well also in extreme cases.
An appropriate identifier encoding promises space gains also in such cases.

Hereafter we shall use the term \cpoi\ (Compact POI) to refer to a compact version of  \poi,
that relies on gapped numeric identifiers
and special encodings.
To grasp the idea
below we explain \cpoi\ over the running example.


\ \\
\noindent
{\bf Example \ref{ex1} (cont.).}
The  lower part of Figure \ref{fig:poi_graph}
shows the gapped representation of the nodes's contents with integers (left),
for our running example.
In this representation
we consider the elements of a node as a list of ids in ascending order and we keep the
first id unchanged while each one of the rest ids is replaced by its difference from the previous id.
The right part of that Figure shows the final form of  \cpoi.
Specifically,
the first value of every node is represented as a normal (32-bit representation) integer,
while the rest values are represented using a special encoding for integers (here, Elias-$\gamma$). \\
$\diamond$

In this paper, apart from proposing \cpoi,
we compare the sizes
of \poi\ and \cpoi\
analytically,
and we identify conditions under which  \cpoi\ guarantees space savings.
Since the conditions are based on bounds
(i.e. they are sufficient, not necessary conditions),
we also report extensive experimental results
over synthetic and real datasets of various characteristics.
We comparatively evaluate
various identifier assignment or reassignment policies,
like
first-in-first-served,
triple frequency,
as well as policies
based on the structure of storage graph of \poi.
The results show that in realistic synthetic datasets,
the compression ratio achieved
by \cpoi\
regarding the contents of the nodes
is 7.5\%.
The only price to pay, in comparison  to the rest approaches is slower additions of new versions,
however since a  version is added only {\em once}, \cpoi\ is a beneficial choice for archiving.

The remainder of this paper is structured as follows:
Section \ref{sec:RW}
discusses related work.
Section \ref{sec:Approach}
introduces \cpoi\
and
provides lower and upper bounds for its space requirements.
Section \ref{sec:Analysis} proposes
methods for assigning identifiers to triples.
Section \ref{sec:Evaluation}
reports extensive comparative experimental results
for various datasets.
Subsequently
Section \ref{sec:Appls}
discusses possible applications of \cpoi,
and finally, Section \ref{sec:Conclusions}
summarizes and concludes the paper.
Proofs
and supplementary measurements
are given at the appendix.

\vspace*{-2mm}

\section{Related Work}
\label{sec:RW}

\begin{figure}
\centering
    \includegraphics[width=40mm]{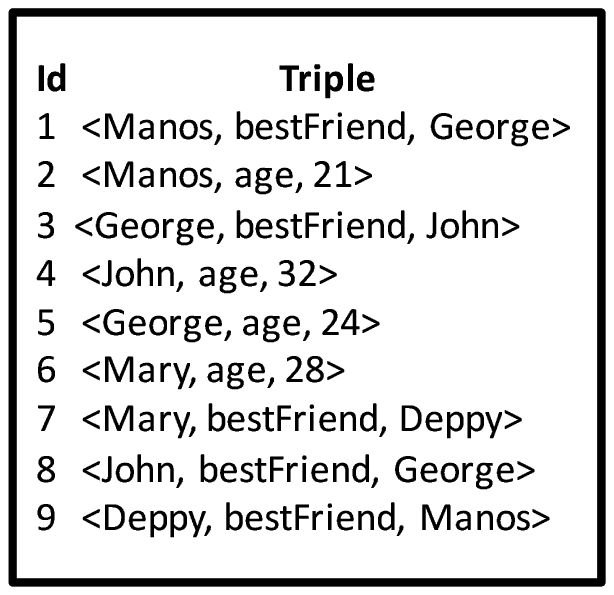}
\vspace*{-0.2cm}
\hspace*{0.9cm}
\caption{Triple strings and their identifiers}
\label{fig:tripleTable}
\vspace*{-0.3cm}
\end{figure}

\begin{figure*}
\centering
    \includegraphics[width=140mm]{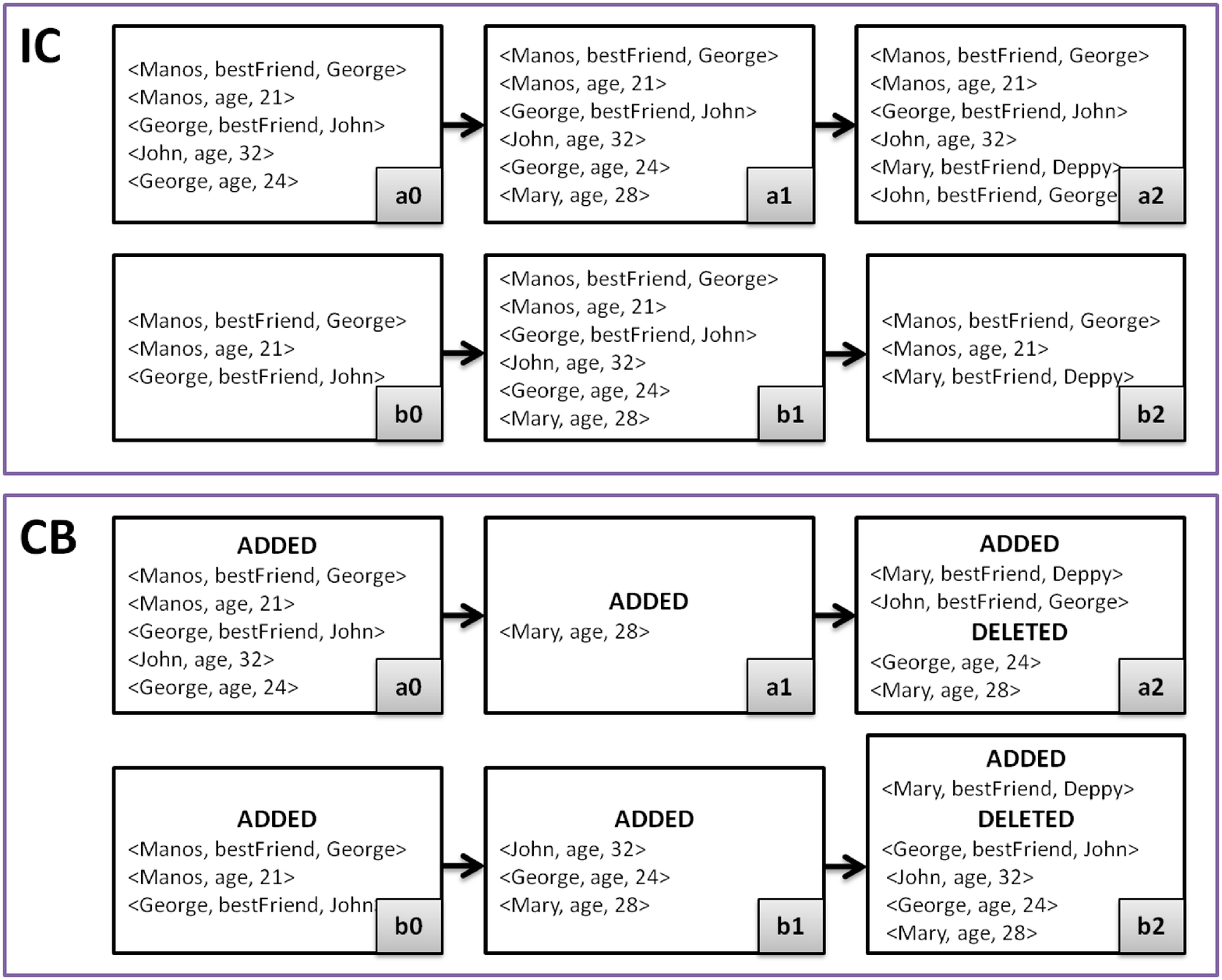}
\vspace*{-0.2cm}
\hspace*{0.9cm}
\caption{Storing versions according to  \ic\ and  \cb\ approach}
\label{fig:IC_CB}
\vspace*{-0.3cm}
\end{figure*}

\begin{figure*}
\centering
    \includegraphics[width=140mm]{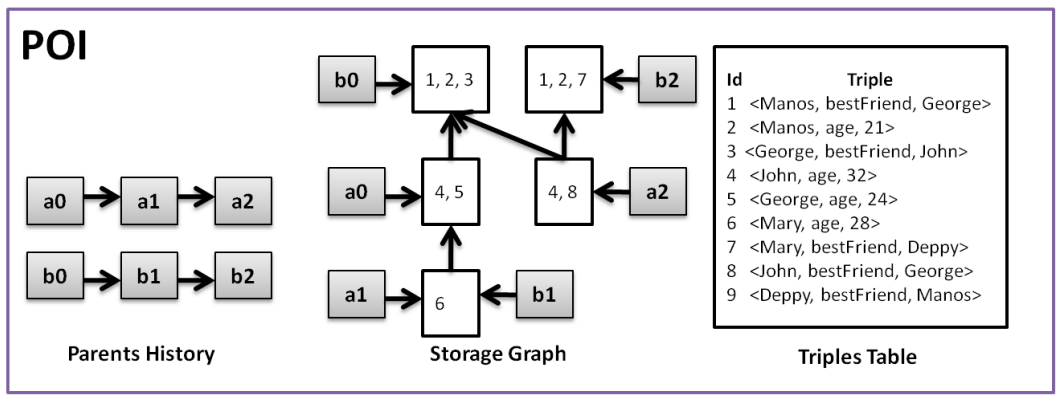}
\vspace*{-0.2cm}
\hspace*{0.9cm}
\caption{Storing versions according to the \poi\ approach}
\label{fig:POI_Structure}
\vspace*{-0.3cm}
\end{figure*}

We shall make clear the differences between
the \ic,  \cb\ and \poi\  approaches
through an example.
Figure \ref{fig:tripleTable}
shows a set of triples and their assigned identifiers.
Figure \ref{fig:IC_CB} illustrates  what
is stored according to the \ic\ and \cb\ approach,
while Figure \ref{fig:POI_Structure}
     shows what will be  stored
according to the \poi\ approach.


\begin{figure*}
\centering
    \includegraphics[width=150mm]{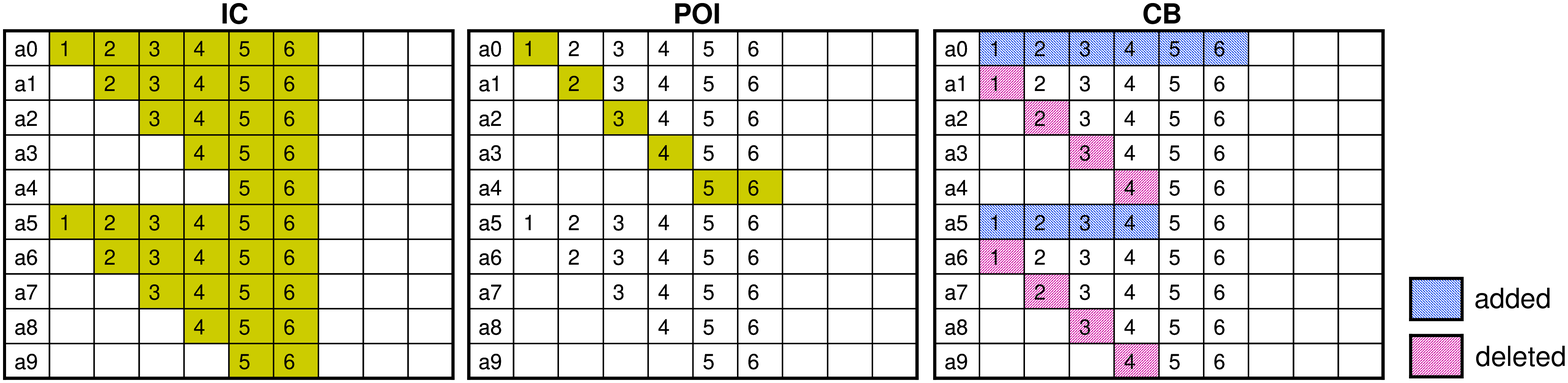}
    \includegraphics[width=150mm]{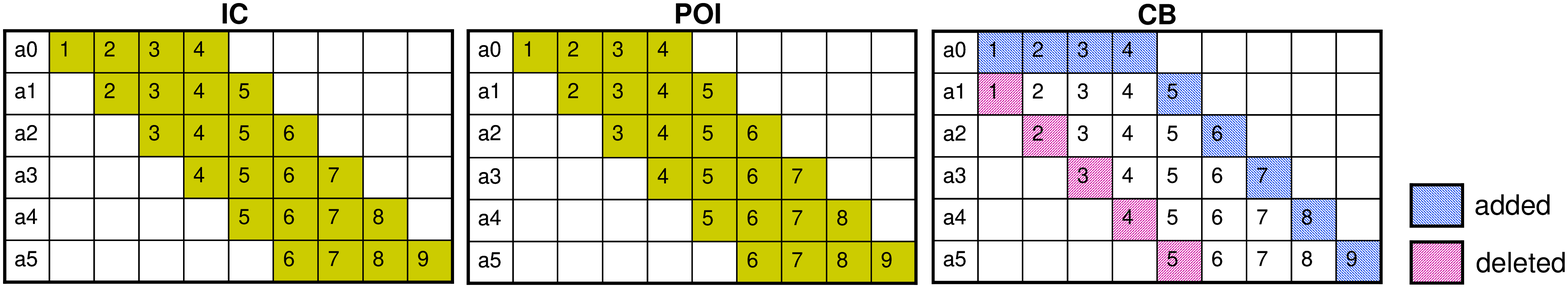}
\vspace*{-0.3cm}
\caption{A good  (upper) and a bad  (bottom) case for \poi}
\label{fig:ExampleTabularBest}
\end{figure*}

Obviously, the
\ic\ approach does not offer any space savings as every version is
stored independently.
\cb\ behaves well
in cases like:
the contents of the KBs form a chain with respect to $\subseteq$ and
they are consecutive in the version history.
The worst case for \cb\ occurs when we have a track
where the same set of
triples is once added and once deleted in an alternative fashion.
In that case \cb\ requires more space than \ic\ (even 2 times worse).
The worst case for \poi\ is when all nodes of the
storage graph are leaves (i.e. the graph is flat),
leading  to space requirements equal to those of \ic.
On the other hand, the best case for POI, is
when the content of every version is a subset of the content of
every version with greater (or equal) content cardinality.
In that
case every triple id is stored only once in the storage graph.
Fig.
\ref{fig:ExampleTabularBest}
illustrates a good and bad case for \poi\
    (this figure illustrates the various approaches
    only  according to triple identifiers,
    neither triple strings nor any other data structure-related cost).

Also note
that the best case for both \cb\ and \poi\ leads to the same space
requirements, while this does not hold for the worst case, which is
much better in the case of \poi.
%

Regarding the \tcb\ (timestamp) approach,
each distinct triple is enriched with pairs of in/out timestamps
(similar in spirit with  \cite{buneman2004asd} for XML).
\tcb\ is not beneficial for version sequences which do not form chains
(as timestamps presuppose a linear order of versions,
e.g. \cite{DBLP:conf/otm/KirstenHGR09} supports only chains).
The reconstruction of the contents of a version is slow in \tcb\ as
it requires looking  up the timestamps of all
triples, unless extra structures (with additional space costs) are adopted.
The only fast task is to check whether a particular triple
belongs to a version, but this is not important  for our application scenario
(version archiving).
Instead \poi\ offers fast reconstruction of version contents
(by taking the union of the nodes which are parents of the pointing node)
and it can save space even in non consecutive versions.

Other approaches,
like \cite{DBLP:journals/pvldb/NeumannW08,iswc/Fernandez10},
aim
at compressing the RDF triples
by using ids for each (subject-predicate-object) element of the triples,
and providing indexes that can be exploited for  query processing.\footnote{
    \cite{DBLP:journals/pvldb/NeumannW10} presents an extension
    of the system in \cite{DBLP:journals/pvldb/NeumannW08}
    that offers versioning services
    which essentially adopts the \tcb\ approach.
}
These approaches are complementary to our work.
\poi\ and \cpoi\  do not compress the set of distinct triple strings,
i.e. the table $TString$.
One can synthesize
\cpoi\ with
these techniques
to further reduce the overall space,
specifically to reduce the space occupied by the table $TStrings$.
We should also make clear that our focus is on archiving, not on
time-travel query services.

We should also note that various {\em cross-version operations},
e.g. containment  checking,
can clearly benefit from a \poi.
Let $v_i$ and $v_j$ denote the contents of two versions,
and $n_i$ and $n_j$ be their corresponding nodes in the
storage graph.
To decide whether a $v_i \subseteq v_j$
one can pose a reachability query on the storage graph,
for checking whether $n_i$ is direct or indirect parent of $n_j$,
thus no need to access the contents of any version.
By adopting a labeling scheme
\cite{JinXRW08}
for the storage graph
we can  decide containment in $\OU(1)$.

Although
\cpoi\
can be considered as a general method to archive
{\em families of sets},
we focus on its use over families of sets of RDF/S triples
since it is a hot application domain that could benefit from  \cpoi.

\section{Introducing and Analyzing CPOI}
\label{sec:Approach}

As mentioned earlier there is a table $TStrings$ which
maps each distinct triple string to a distinct numeric identifier.
Let $T$ be the set of these identifiers,
and $|T|$ denote the cardinality of this set.


The \emph{storage graph} $\Upsilon$ of a \poi\ is a pair
$\langle \Gamma, stored\rangle$ where $\Gamma = (N,R)$ is a directed acyclic graph
and $stored$ is a function from the set of nodes $N$ to the powerset of
$T$.
%

Each node $n$ of the storage graph of \poi\ holds a set of
numeric triple identifiers,
denoted by $n.stored$
(where $n.stored \subseteq T$).

We can keep this set as a list sorted in ascending order.
This list can be considered as a
sequence of gaps between triples identifiers,
e.g. for the sequence
[32011, 32013, 32014, 32017],
the sequence of gaps would be [32011, 2, 1, 3]
(of course
the original identifiers can  be recomputed through sums over the gaps).
This \emph{d-gapped} representation as it is commonly known,
is very popular in the area of IR \cite{WittenMoffatBell99}.

The  list of gapped identifiers can then be compressed using a suitable
compression scheme.
Compression is obtained by encoding small values with shorter codes.
Special encodings,
such as
Elias-$\gamma$,
$\delta$ \cite{Elias},
Golomb-Rice \cite{Golomb},
Binary Interpolative Coding \cite{593998}
and Variable-byte encoding,
are efficient for small integers, as the smaller the integer is the less space is needed.
For example,
 a positive integer $k$,
can be represented by
$2\lfloor \log_{2}k \rfloor +1$ bits in Elias-$\gamma$.
However, when integers become large, the storage space also becomes large.
Consider the list [32011, 2, 1, 3].
Since we expect to have small numbers only in gaps,
while the first number can be any number,
it is beneficial to encode the first number
with a fixed length encoding, e.g. by 4 bytes,
and the rest
with a variable length encoding (e.g. Elias-$\gamma$).
This means that the original  list [32011, 32013, 32014, 32017]
is actually represented by
the following list of bit sequences
$[00000000000000000111110100001011, 010, 1, 011]$.
It follows that the more id numbers are close to each other in every
node, the more space saving can be achieved.

\vspace*{-2mm}

\subsubsection*{The Gaps of a Node}

To simplify notations,
we shall  use the same symbol $n$ to
denote a node and
its set of identifiers, i.e. $n.stored$.
Consequently $|n|$ will refer to the cardinality of this set.
Clearly,
the space  required by a node
depends on both $|n|$ and the way we represent
the ids of the triples  in  $n$.
Regarding the latter let us define the sum of
gaps between consecutive ids, as this determines (approximates)
the total size of ids (and it is independent of any particular encoding scheme).
Indeed, such sum of gaps coincides with the bits required if
the \emph{unary}
representation\footnote{
    Each positive integer $k$ is represented by \emph{k} bits,
    specifically by \emph{k} - 1 ones followed by one zero
    (or \emph{k} - 1 zeros followed by an one).
}
is used.
Specifically, if  $n$ = $\{tr_{1}, ... , tr_{|n|}\}$, we define:
\vspace*{-1mm}
\begin{equation} \label{eq:gaps1}
 gaps(n) = \sum_{i=1}^{|n|-1}{(tr_{i+1}-tr_{i})}
\end{equation}
For example, if $n$ = $\{1, 4, 8\}$, then $gaps(n)$ = 3+4 =
7.  The smaller the value of $gaps(n)$ is, the better representation
(for $n$) can be achieved. It is not hard to see that:
$ |n|-1 \leq gaps(n) \leq |T|-1$.
The minimum value and therefore the best
case for $gaps(n)$ is  achieved when the ids are consecutive numbers,
e.g. $n$ = $\{8, 9, 10\}$. On the other hand, the worst
case for $gaps(n)$ is when $n$ contains ids that cover
the entire range of values (hence from 1 to $|T|$). In that case,
$gaps(n)$ equals to $|T|-1$,
as it actually expresses  the
transition from 1 to $|T|$ and each id covers one step of that transition.
For example, if $T$ = $\{1, ..., 100\}$, the worst representation for
a node $n$ such that $|n|$=3, leads to $gaps(n)$ = 99.
e.g. both
$n$ = $\{1, 5, 100\}$ and $n$ = $\{1, 80, 100\}$
lead to the same value for $gaps(n)$.
It follows from this observation
that if we know the id of the first and the last element
of $n$,
then we can compute $gaps(n)$ without having to use formula (\ref{eq:gaps1}), since it holds:
\vspace*{-2mm}
\begin{equation} \label{eq:gaps}
 gaps(n) = \sum_{i=1}^{|n|-1}{(tr_{i+1}-tr_{i})} =  tr_{|n|}-tr_{1}
\end{equation}


We can define the total gaps of all nodes of the storage graph $\Gamma = (N,R)$ as:
$ Gaps(N) = \sum_{n\in N}{gaps(n)}$,
where
 $N = \{n_{1}, ... , n_{|N|}\}$ is a family of subsets of $T$.
Below we identify lower and upper bound for $Gaps(N)$.

\vspace*{-2mm}

\begin{myproposition} \label{prop:AllDisjoint} {\em
If a storage graph has $|N|$ sets
and they are \emph{pairwise disjoint}, then:
$
 |T|-|N| \leq Gaps(N) \leq |N|(|T|-|N|)
$.
$\diamond$}
\end{myproposition}

\vspace*{-2mm}

Note that if all sets of $N$ are pairwise disjoint
then it always holds that
$|T| - |N| \geq 0$,
i.e.
$|T| \geq |N|$,
since the maximum number of sets in a partition of a set $s$,
is equal to $|s|$
(that partition consists of singletons).
In the extreme case where
$|T| = |N|$
(and thus all nodes are singletons)
the gaps are indeed 0.

Regarding the general case,
where we can have overlaps,
it holds:

\vspace*{-2mm}

\begin{myproposition} \label{prop:Overlaps} {\em
If a storage graph has $|N|$ sets
and there are \emph{overlaps} then:
$
\sum_{i=1}^{|N|}|n_{i}|-|N| \leq Gaps(N) \leq |N||T|-|N|
$.
$\diamond$}
\end{myproposition}

\vspace*{-2mm}

\vspace*{-2mm}

\subsubsection*{Gaps and Storage Space}

If we consider that the first id of every node is represented by a fixed length encoding
of $\B$ bits (usually 32),
and the successive (gapped) ids are coded using \emph{unary} codes,
then the storage space required, measured in bits, and denoted by $Space_{CPOI}(N)$, is:
\begin{equation} \label{eq:space_cpoi}
Space_{CPOI}(N) =  \B*|N| + Gaps(N)
\end{equation}
Without a gapped and unary encoded representation, the required space by a $\B$-bit representation is:
\vspace*{-2mm}
\begin{equation} \label{eq:space_poi}
Space_{POI}(N) =  \B*\sum_{i=1}^{|N|}|n_{i}|
\end{equation}

\vspace*{-4mm}

\subsubsection*{Uniform Codes for Ids}

Instead of using $\B$ bits per integer,
or adopting a gapped and specially encoded representation,
we could use a {\em uniform representation} of
$\lceil \log_{2}|T| \rceil$ bits for each id.
Obviously this leads to  space savings. For example,
if $|T|=10^{6}$ then instead of $\B$ bits, we could use
$\log_{2}10^{6} = 20$ bits.
Hence,
for $\B=32$
we can achieve compression ratio of
$\frac{20*\sum_{i=1}^{|N|}|n_{i}|}{32*\sum_{i=1}^{|N|}|n_{i}|}100\%
 \simeq 60\%$,
i.e. the compressed space is around $60\%$   of the original space.
Hereafter, we will denote the space required by the above representation as
$Space_{CPOI_U}(N)$, where the `U' comes from \emph{uniform}.
It follows that:
\vspace*{-3mm}
\begin{equation} \label{eq:cpoiu}
Space_{CPOI_U}(N) = \sum_{i=1}^{|N|}|n_{i}| * \lceil \log_{2}|T| \rceil
\end{equation}
Note that in uniform representation the number of required bits is definite for
a specific $T$, while in unary representation that depends
on the assignment of triples' ids,
and on the value of  $Gaps(N)$.
However we should note
that the uniform representation
is not practical for the problem at hand
due to the limitation of the
number of integers that it can encode.
The  insertion of new versions with brand-new
triples would require changing all triple identifiers
and consequently the contents of all nodes.

Table \ref{tbl:Codings} summarizes the occupied space of each approach.

\begin{table}
\footnotesize{
\centering
\begin{tabular}{|l|l|}
\hline
{\bf  Encoding} & {\bf Storage Space for Nodes} \\\hline\hline
{\bf  \poi}    & $\B*\sum_{i=1}^{|N|}|n_{i}|$ \\\hline
{\bf  \cpoi}     & $\B*|N| + Gaps(N)$ \\\hline
{\bf  \cpoiu}     & $\sum_{i=1}^{|N|}|n_{i}| * \lceil \log_{2}|T| \rceil$ \\\hline
\end{tabular}
\caption{Space (in bits) by the nodes of each  method}
\label{tbl:Codings}
}
\vspace*{-0.3cm}
\end{table}

Below we compare analytically the above approaches.
\cpoi\ requires less space than \poi\ if
 $Space_{CPOI}(N) < Space_{POI}(N)$, i.e.
 if
$ Gaps(N) < \B*\sum_{i=1}^{|N|}(|n_{i}|-1)$.

\begin{myproposition} \label{prop:Gain} {\em
If $Gaps(N)$ is at most
$\B$ times more than the
lower bound of $Gaps(N)$,
i.e. if
$Gaps(N) < \B*\sum_{i=1}^{|N|}(|n_{i}|-1)$,
then \cpoi\  requires less space than \poi.
$\diamond$
}
\end{myproposition}

By combining Prop. \ref{prop:Overlaps} and Prop. \ref{prop:Gain}
we get:

\begin{myproposition} \label{prop:Avg} {\em
 If  the number of distinct triples (i.e. $|T|$)
 is not greater than $\B$ times the average number of elements of a node,
 then \cpoi\
 requires less space than \poi.
$\diamond$
}
\end{myproposition}

Note that since  Prop.
\ref{prop:Avg}
is based on the  extreme case where
we have the worst case for $Gaps(N)$,
it specifies  a sufficient (not necessary) condition.
This means that we can still gain with a \cpoi\
even if the condition of
Prop. \ref{prop:Avg} does not hold.

Regarding the uniform representation, we have:

\begin{myproposition} \label{prop:Uniform}{\em
\cpoi\ requires less space than \cpoi$_U$, iff $Gaps(N) \leq
\sum_{i=1}^{|N|}|n_{i}| * \lceil \log_{2}|T| \rceil - \B*|N|$.
$\diamond$ }
\end{myproposition}


We can compare \cpoi\ and \cpoiu\ also wrt the worst/best case of \cpoi\
(as expressed in Prop. \ref{prop:Overlaps} and equations (\ref{eq:space_cpoi}) and (\ref{eq:cpoiu})),
i.e. in a way that does not require knowledge of $Gaps(N)$.

\vspace*{-1mm}

\begin{myproposition} \label{prop:WUBUE} {\em
The worst case of unary is better than uniform encoding, when:
\vspace*{-3mm}
\begin{equation} \label{eq:cpoiAlwaysBetterThanCpoiu}
avg(|n_i|) * \lceil \log_{2}|T| \rceil \geq |T|+\B-1
\end{equation}
}
\end{myproposition}
\vspace*{-1mm}
It follows that unary representation is better than uniform when we have
high average node size and small $|T|$,
i.e. large overlapping.

The other way around:
\vspace*{-1mm}
\begin{myproposition} \label{prop:BUWUE} {\em
The best case of unary is worse than uniform encoding,
and therefore
uniform is certainly better than unary representation,
when:
\vspace*{-2mm}
\begin{equation} \label{eq:cpoiAlwaysWorstThanCpoiu}
avg(|n_i|) * (\lceil \log_{2}|T| \rceil -1) \leq \B-1
\end{equation}
}
\end{myproposition}


\begin{figure}
\centering
    \includegraphics[width=80mm]{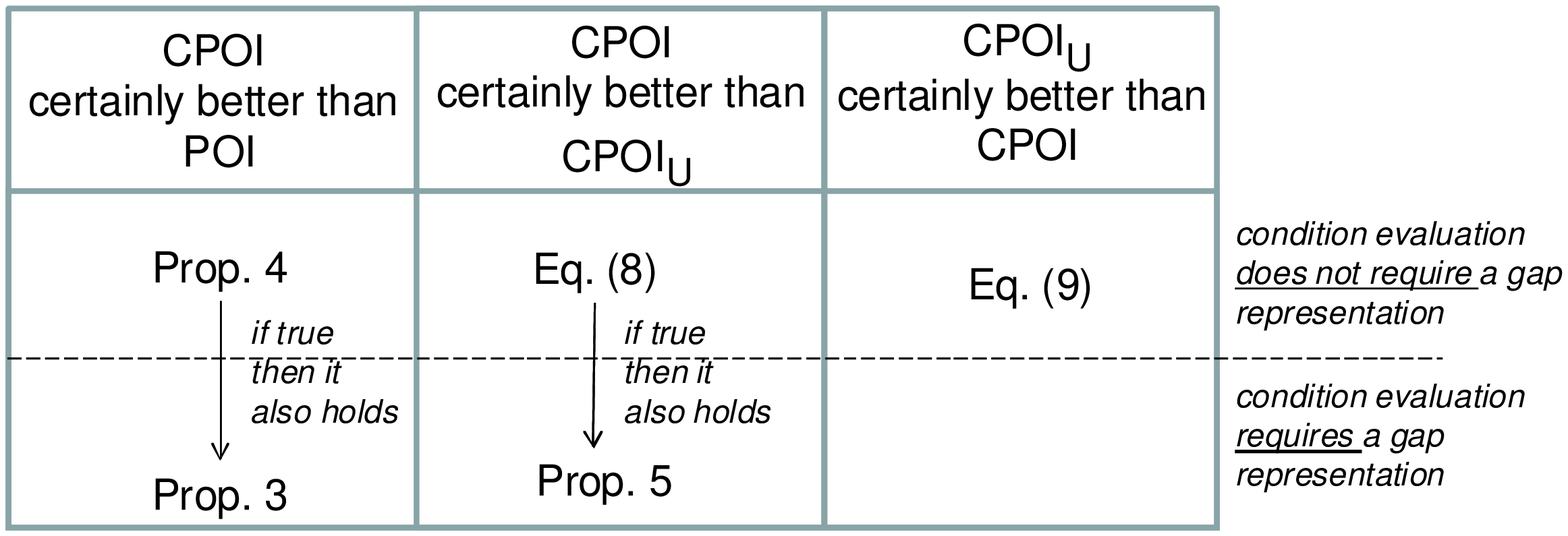}
\vspace*{-0.2cm}
\hspace*{0.9cm}
\caption{Synopsis of analytical results}
\label{fig:analyticalResults}
\vspace*{-0.3cm}
\end{figure}

\vspace*{-2mm}

\noindent
{\bf Synopsis}.
Figure \ref{fig:analyticalResults} illustrates the conditions under which
one  choice is certainly better (requires less space) than another.
Notice that we can check whether some conditions hold or not
(specifically those in
Prop. \ref{prop:Avg} and equations
(\ref{eq:cpoiAlwaysBetterThanCpoiu}) and (\ref{eq:cpoiAlwaysWorstThanCpoiu})),
even if we have a plain \poi, i.e. without the need of a gapped representation.
On the other hand the conditions that refer to $Gaps(N)$
require having a gapped representation
and the value of $Gaps(N)$ depends on the way identifiers have been assigned.
However, we can have space savings even if the above conditions are not met,
since the conditions rely on lower and upper bounds. 
This motivates the experimental evaluation of Section \ref{sec:Evaluation},
where we attempt to compare
assignment approaches  (aiming at approaching the lower bound of $Gaps(N)$),
and investigate the amount of space saving that we can achieve.



\section{Assignment Policies for Identifiers}
\label{sec:Analysis}

Our objective here is to investigate policies
for assigning  ids
aiming at achieving a small  value for  $Gaps(N)$.
We need a method that takes as input
the storage graph of \poi\
and
reassigns the ids of triples
so that each node to contain ids close to each other.
Moreover,
that method should be also efficient in time.
Below we discuss a number of approaches:

\vspace*{0.2cm}
\noindent
$\bullet$ {\bf Default (id assigned at the first appearance of triple)}.
According to this policy,
each triple is assigned an id,
the first time that it appears in one version,
and it gets the smallest available integer.
Obviously, this assignment does not depend on the structure of \poi.

\vspace*{0.2cm}
\noindent
$\bullet$ {\bf Node Size.}
We  order the nodes of the storage graph by their \texttt{size}
(i.e. $|n_i|$) and then we assign ids to their triples
starting either from the larger nodes,
or from the smaller nodes.
If we start from the larger nodes (\emph{Size})
then these nodes are favored,
so we can expect gains for them.
An alternative approach is to start assigning identifiers
starting from the smaller nodes (\emph{Size\_Rev}).
The motivation
is that a small node can ``waste" more bits per id,
than a large node in a ``bad" assignment.
Specifically, consider two nodes $n_i$ and $n_j$ such that $|n_i| < |n_j|$.
We can define the ``bits per id of a node $n$" as follows:
$bpi(n) = \frac{|gaps(n)|}{|n|}$.
The worst (i.e. space consuming)
case is $bpi(n) = \frac{|T|-1}{|n|}$.
It follows that
$worst(bpi(n_j)) < worst(bpi(n_i))$,
since  $\frac{|T|-1}{|n_j|} <  \frac{|T|-1}{|n_i|}$.
By starting from the small nodes,
we expect a better assignment for them,
an assignment that leads to smaller  $bpi(\cdot)$ values.
%
Obviously, nodes with size equal to 1,
do not benefit from a gapped representation,
therefore we assign identifiers to their triples
at the end.
In conclusion, each policy (from large or small nodes) has its own pros and cons.


\vspace*{0.2cm}
\noindent
$\bullet$ {\bf Triple Frequency.}
For each triple in $T$ we count the number of nodes
that contain it
and then we order the triples according
to this number,
getting a list of the form:
$\langle$ appearing $k$ times $\rangle$,
$\langle$ appearing $k-1$ times $\rangle$, $\ldots$ , $\langle$
appearing 1 time $\rangle$.
Then we assign ids starting from the most frequently occurring triples,
aiming at achieving consecutive (or close) ids
in several nodes.
If we want to reduce the maximum gap between an id and the rest ids,
then it is beneficial to assign to that id the value $|T|/2$,
since the max gap in that case is $|T|/2$.
The ids that  can give the maximum gaps (with the rest),
are those in the ends of the interval,
i.e. $1$ and $|T|$.
So one reasonable approach would be to start giving ids to the
frequent triples starting by $|T|/2$
and then continue using ids based on their  absolute distance from
$|T|/2$ (e.g. if $|T|/2=50$, then consume ids
in the following order 50,51,49,52,48, and so on).
It follows that the least occurring triples
will get ids close to 1 or $|T|$.
%
%
Moreover, note that a triple with $f$=1 does not affect other nodes
than the one it appears in,
so we do not have to be concerned about
the ids of triples with $f=1$
(i.e. no overlapping issues as they appear only in one node).
Furthermore, we can exploit the fact that the triples
in each frequency list are grouped so that those of the same node are
adjacent, for giving them consecutive ids.
Finally, if we consider that the distribution of the triples in nodes
follows a power-law,
then the expected size of the list with
$f$=1 is much bigger than the sizes of the other lists.
Considering all the above, we start the assignment from the list of triples
with $f$=1 and $id=1$.
When we have consumed the half of that list,
we continue the assignment with the most frequently occurring triples. The rest lists
follow (according to their frequency) and at the end we assign the
remaining half of triples  with $f$=1.\footnote{
    An example is given at the end of this section.
}
%
%
%
This reassignment is more expensive than the previous ones since
it requires computing  the frequency of all triples.

\vspace*{0.2cm}
\noindent
$\bullet$ {\bf Storage Graph.}
We can traverse the  storage graph
and assign ids by the order  we encounter nodes.
Let's first make some remarks regarding the storage graph:
(a) If two  nodes are connected (i.e. one is a direct or indirect child of the other)
then they certainly have disjoint content.
This means that overlaps can occur between nodes which  are not connected
(and clearly nodes of the same level  fall into this category).
(b) Each node of the storage graph is pointed to by at least one version.
Since the contents of the versions that a node represents
equals the triples  stored at that node plus
the triple stores in all parent nodes of that node,
it follows that
if the variance of the version contents' sizes is small,
then the maximal nodes of the storage graph
(i.e. those which have not any parent),
%
are expected to store more triples
than the deeper nodes.
Based on the above observations,
one approach is to traverse the
storage graph in a \texttt{Breadth-First Search (BFS)} manner,
i.e.  to  start from high level nodes and then to descend.
In this way it is expected that we will encounter larger nodes
at the beginning and such nodes can (is not impossible to) overlap.
Alternatively, and with the same motivation
with the policy {\em Size\_Rev},
we could adopt a {\em reverse} BFS policy ({\em BFS\_Rev}).
In comparison to {\em Size} policy,
the storage graph-based policies have the following benefit:
successive (during the assignment) nodes have higher probability to have overlaps.


\ \\
With respect to computational cost,
apart from {\em Frequency}, the rest reassignment policies
do not require any preprocessing,
and they are very fast (roughly each method requires traversing the storage graph once).

\begin{figure}
\centering
    \includegraphics[width=3.4in]{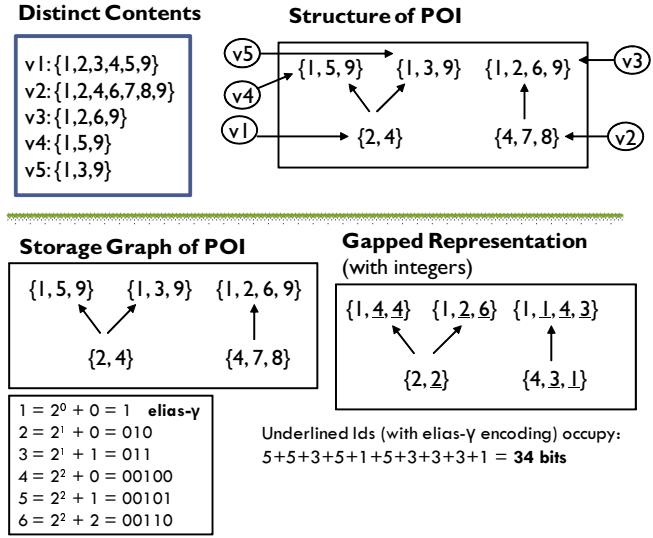}
\vspace*{-0.2cm}
\caption{Storage graph of \poi\ without and with a gapped representation}
\label{fig:exampleReassignment}
\end{figure}

\begin{figure*}
\centering
    \includegraphics[width=2.785in]{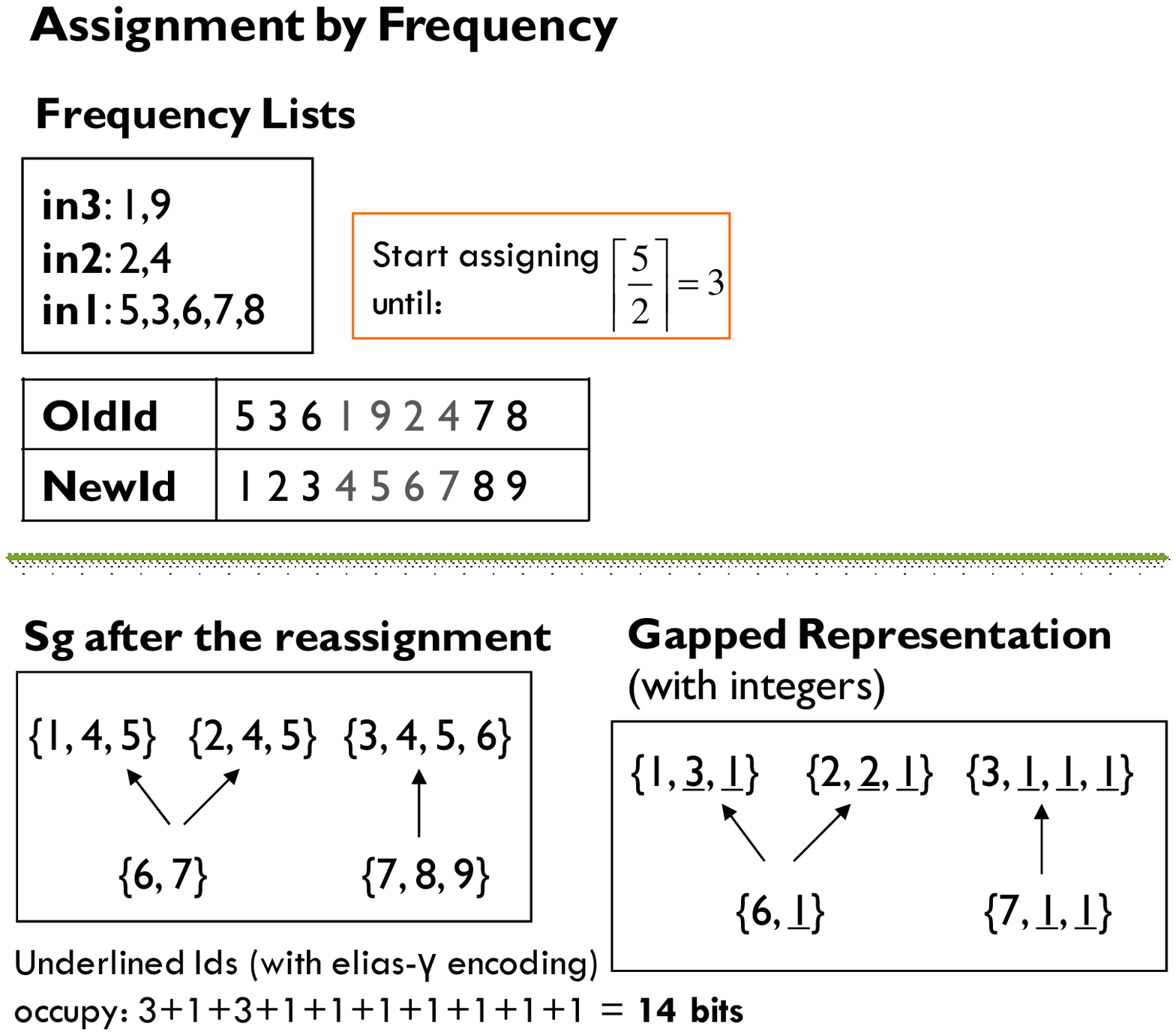}
    \includegraphics[width=2.785in]{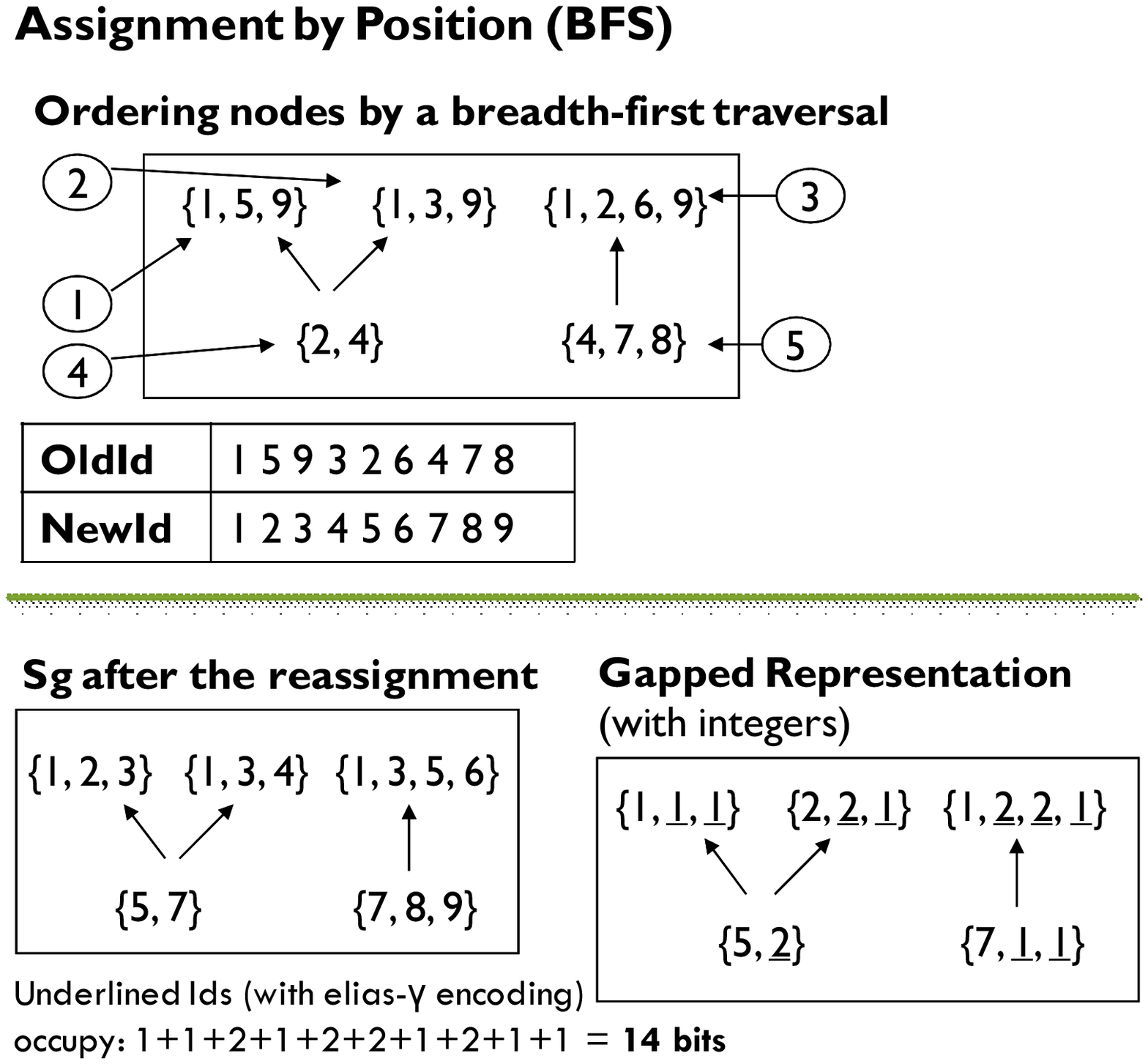}
\vspace*{-0.2cm}
\caption{Assignment by Frequency (left) and by BFS (right)}
\label{fig:exampleReassignment1}
\end{figure*}

\begin{example}
Here we explain two of the assignment approaches
({\em Frequency} and {\em BFS}) through an example.
Figure \ref{fig:exampleReassignment} shows the distinct contents of
5 versions and the corresponding graph of \poi\ before and after the
adoption of a gapped representation.
Now the left part of Figure \ref{fig:exampleReassignment1} (resp. right)
depicts the procedure of reassignment by {\em Frequency} (resp. by {\em BFS}).
Regarding the former, we first create the frequency lists and then
we start the assignment from the list with frequency=1
until the half of that list (in our example we will consume three triples).
We continue the assignment with the most frequent triples
(i.e. triples appearing in three nodes,
in two nodes and finally the remaining two
triples of the first list).
Subsequently, we update our storage graph
with the new ids.
The observe that the resulting gapped representation needs less space
than the initial graph,
specifically even in this small example
the space is almost 2.5 times less.
\end{example}

\section{Experimental Evaluation}\label{sec:Evaluation}



To evaluate the above policies in large datasets we generated and used
{\em synthetic} datasets.
We have to note that related work mainly reports results over synthetic datasets
or over real datasets which are not versioned.
For these reasons we decided to derive
and use three kinds of {\em synthetic} datasets.

\subsection{Datasets}

\begin{figure}
\centering
    \includegraphics[width=61mm]{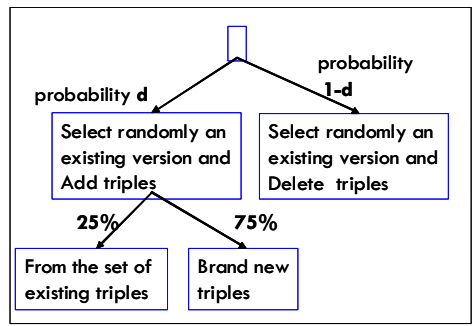}
\vspace*{-0.3cm}
\caption{Synthetic dataset generation method for \DatA }
\label{fig:DataSetGenerationInitial}
\vspace*{-0.3cm}
\end{figure}

\godown
{\bf Dat1}.
Using the synthetic KB generator described in
\cite{DBLP:conf/esws/TzitzikasTA08}, we created a dataset (\DatA)
consisting of  1000 versions, each having 10,000 triples on average,
where the size of each triple is 100 bytes (a typical triple size).
The version generation method is described next
and  illustrated in Figure \ref{fig:DataSetGenerationInitial}.
As in real case scenarios, a new version is commonly  produced by modifying
an existing version.
In order to generate the content of a new version,
we first choose at random a parent version and then we either \emph{add} or
\emph{delete} triples from the parent contents. The difference in triples
with respect to the parent content is 10\%, i.e. 1000 triples. We have an
additional parameter $d$ that defines the probability to choose triple
additions (so with probability $1 - d$ we delete triples).
In this respect, we create versions whose contents are either supersets or
subsets of the contents of existing versions. We experimented with $d$ in the
range of [0.5, 0.9] (we ignored values smaller than 0.5 as deletions usually
do not exceed additions). For additions, we assumed that the $25\%$ of the
additional triples are triples which already exist in the KB (in the content of
a different than the parent version), while the rest $75\%$ are brand new triples.
This is motivated by the fact that in a versioning system it is more rare
to re-add a triple which exists in an old version and was removed in one of the
subsequent versions, than to add new triples.
Notice that as $d$ increases, more new triples are created and less are deleted
(so the total number of distinct triples increases).

\begin{figure}
\centering
\includegraphics[width=73mm]{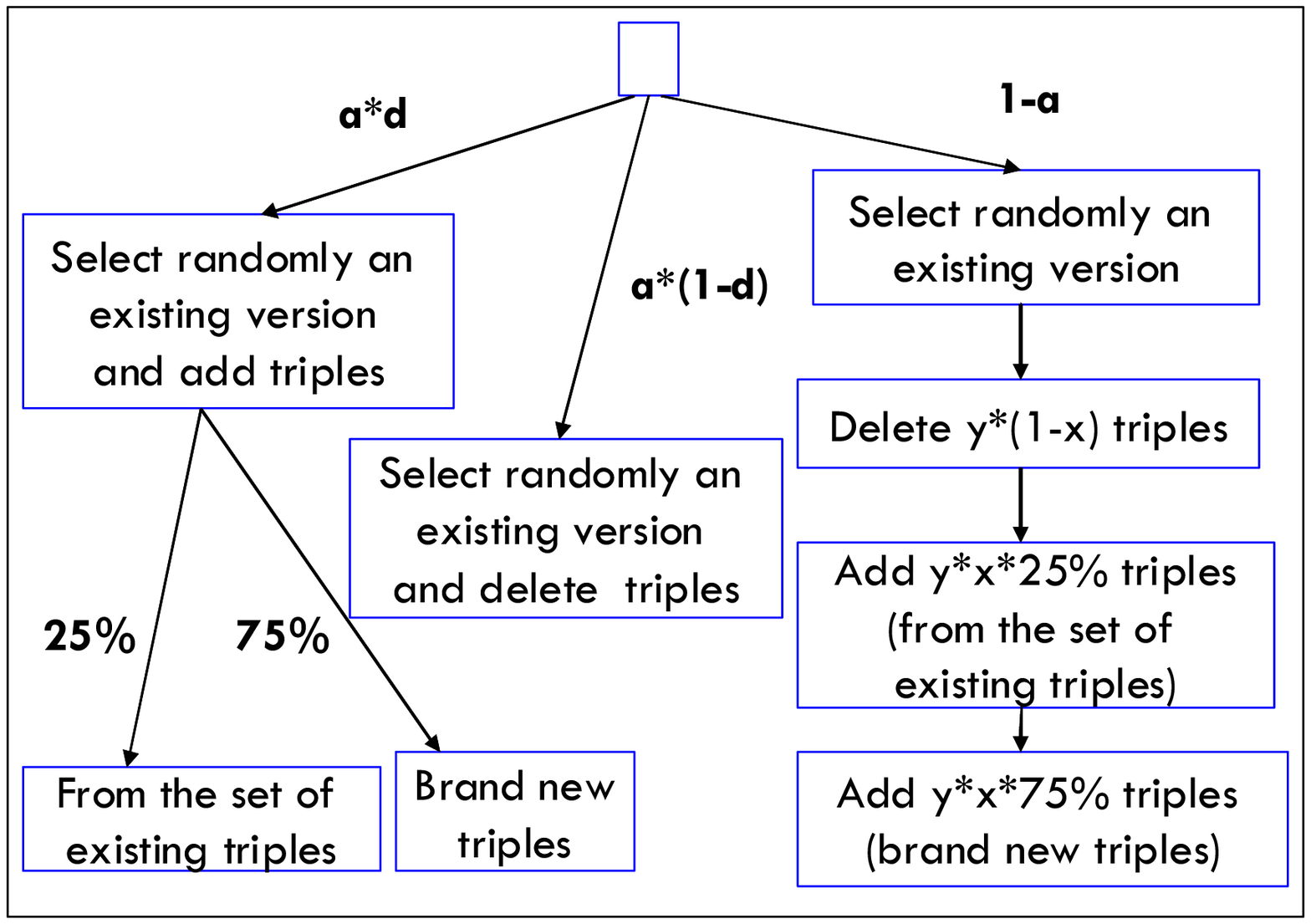}
\vspace*{-0.3cm}
\caption{Synthetic dataset generation method for \DatB}
\label{fig:DataSetGeneration}
\vspace*{-0.3cm}
\end{figure}

\godown
{\bf Dat2}.
The \DatA\ consists of  versions
whose contents are proper subset or superset of the
contents of the parent versions.
To obtain a more
realistic dataset,
containing less $\subset$ relationships over contents,
we created another dataset  (\DatB)
with a different version generation method
illustrated in Figure \ref{fig:DataSetGeneration}.
At first, we choose at
random an existing version as a parent and then we either only add or delete triples from
the parent content, or we make \emph{both} triple additions and deletions. In the first case,
we create versions whose contents are either subsets or supersets of the contents of
existing versions, like \DatA. We have an additional parameter $a$ that defines the
probability to choose the first case (and $1-a$ the second case). We experimented with $a$=0.3
(so the probability a version to be proper subset or superset of the parent version is 0.3).
In the second case, we add and delete triples from the parent content. We use parameter $x$ to
specify the proportion of added triples (and $1-x$ for the deleted triples).
The parameter $x$
ranges [0.5, 0.9], so $1-x$ ranges (0.1, 0.5).
We used $x=d$ for the production of the dataset used in the experiments.
The difference in triples with respect to the
parent content is $10\%$
This proportion is represented by the parameter $y$.
For additions, in both cases
(i.e. $a$ and $1-a$ cases)
we assumed that the $25\%$ of the additional triples are triples which already
exist in the KB, while the rest $75\%$ are brand new triples.
\DatB\  consists of 1000 versions, each having 10,000 triples on
average as well.

\begin{figure}
    \centering
        \includegraphics[width=60mm]{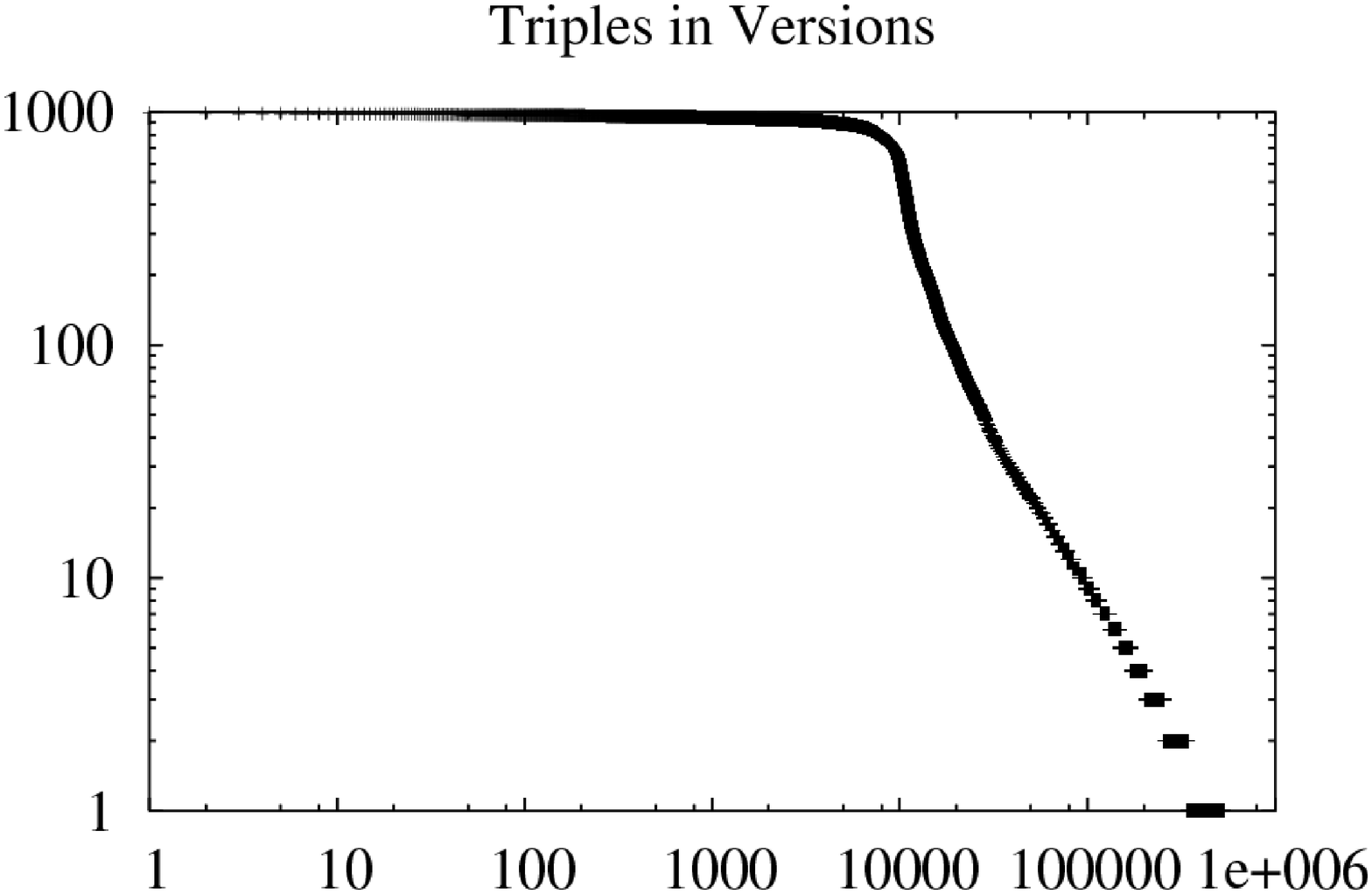}
        \includegraphics[width=60mm]{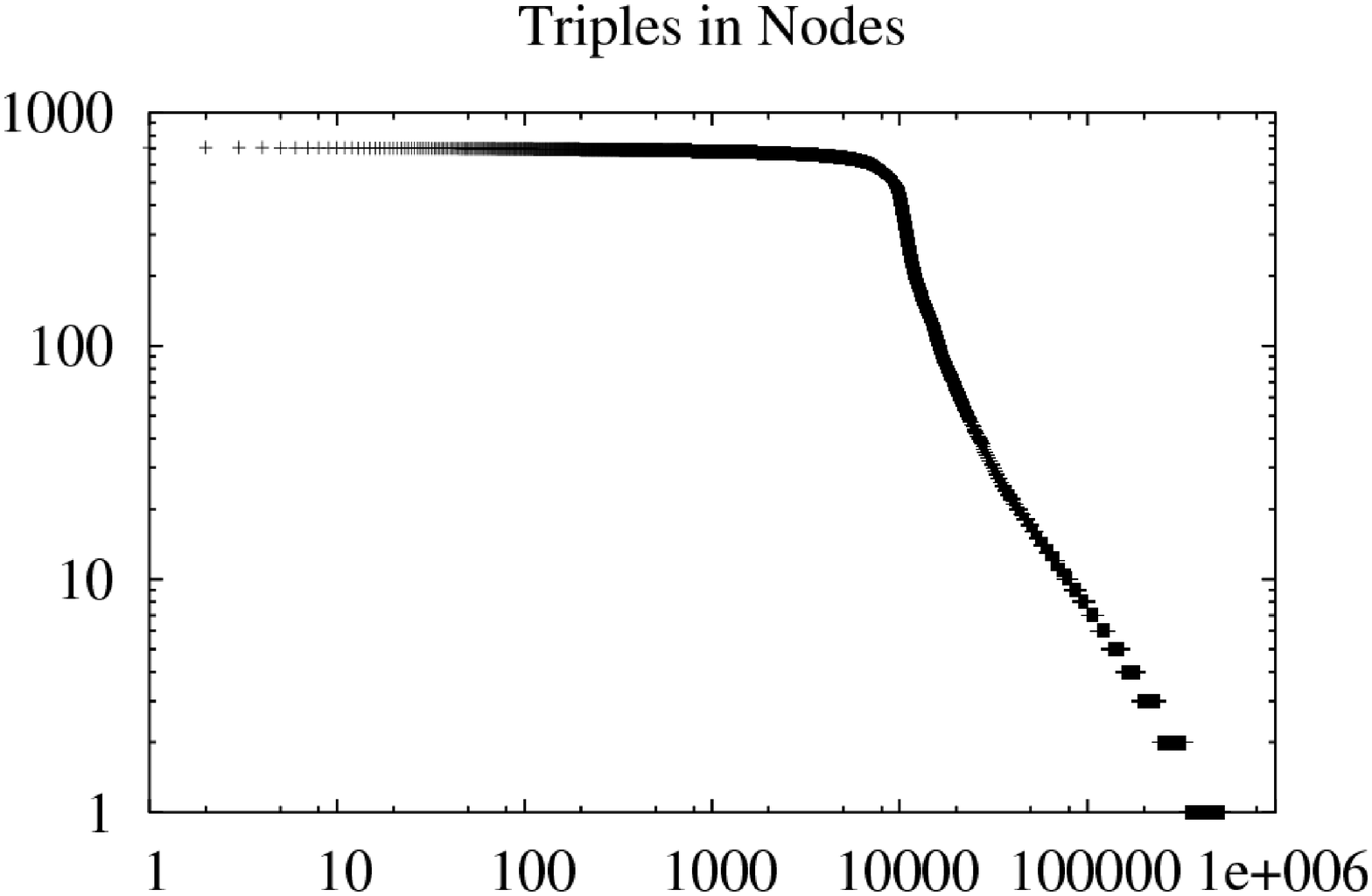}
    \caption{The distribution (in logscale) of triples in versions (up)
             and in nodes (down) for \DatB\ for d=0.7 (Y: frequency, X: triples)}
    \vspace*{-0.3cm}
    \label{fig:powerLaw}
    \end{figure}

The frequency of triples in versions
approximates a power law distribution.
This hold for both the frequency of triples
in versions, and the frequency of triples
in the nodes of \poi.
%
Just indicatively,  Figure \ref{fig:powerLaw}
shows in log scale the distribution of triples
in versions (left), and in nodes (right)
for \DatB\ with $d=0.7$ where $|T|=531,500$.
Notice that if we exclude the first 10,000 triples
the rest 521,500 triples
follow a power-law distribution
as their plot in the log scale
approximates a straight line.

After feeding these datasets to \poi\
we observed that the storage graph of \poi\ for  \DatA\
has a large number of edges, while the storage graph for  \DatB\
has a more flat morphology with less connections and depth.
Specifically, the average depth of the graph for \DatA\ (resp. \DatB)
in our experiments (and assuming all $d$ values)
is 3 - 5.5 (resp. 1.4),
while the max depth for \DatA\ (resp. \DatB) is 12 - 14 (resp. 5).

\godown
{\bf Dat3}.
This dataset consists of 1000 versions, each having 10,000 triples
on average, and $|T|$=400,000.
The key characteristics are:
no version is subset of another,
versions share a lot of common triples and the triple frequency follows a power-law
for being close to real datasets
(the construction method is detailed in the next paragraph).
The dataset is ideal
for
testing
the  worst case for \poi.
Since no version is subset or superset of any other version, the graph of \cpoi\
has a flat structure, i.e. it is like an inverted index
where nodes point to compacted lists of triple ids.
The only space saving offered by \cpoi\ is that whenever two versions have
the same content,  only one node and one posting list is kept in \cpoi.

The versions of \DatC\ were produced  as follows:
at first we compute the frequency
of each distinct triple
using  the formula
    $Freq(tripleId)=(\frac{100.230}{tripleId})^{0,6}+25$,
meaning that each triple appears to at least  25 versions,
creating a list of the form $<$tripleId, \#versions$>$.
Then we start filling each version's content consuming the ids of every list consecutively.
Specifically, once we have inserted one triple in \#versions (according to the list),
we continue with the next triple starting insertions from the subsequent version of
the one that we added the previous tripleId last.


\godown
{\bf GO}.
We also conducted experiments over only a few versions of GO (Gene Ontology).
We used the RDF/S dumps from the GO project.\footnote{www.geneontology.org/}
This dataset contains 27,640 classes, and 1,359 property instances and uses 126 properties to
describe genes. We used only a few versions, specifically 6 versions
(v.16-2-2008, v.25-11-2008, v.24-3-2009, v.5-5-2009, v.26-5-2009, v.22-9-2009)
which {\em are not successive}, so a lot of changes exist between them,
and indeed none of them is subset of another. We have to note that if we had used
a higher number of versions, then that would be an advantage for  \cpoi.

\subsection{Compared Options}

For each dataset we compared the following options:
\begin{compactitem}
\item[a)] plain \poi\ (i.e. no reassignment, 32-bit integer encoding),
\item[b)] \cpoi\ \emph{Default} (i.e. no reassignment),
\item[c)] \cpoi\ after ordering triples randomly (\emph{Random}),
\item[d)] \cpoi\ after ordering triples wrt node size (\emph{Size} in descending order and \emph{Size\_Rev} in ascending order),
\item[e)] \cpoi\ after ordering triples wrt their frequency in nodes (\emph{Frequency}),
\item[f)] \cpoi\ after ordering triples wrt their position in the storage graph (\emph{BFS}
and \emph{BFS\_Rev} for reverse traversal),
\item[g)] \cpoiu\ using uniform encoding. \\
\end{compactitem}
\vspace*{-1.5mm}
We compared the above options with respect to the following aspects:
storage space for the node contents,
and time to assign the identifiers.
For \cpoi\ we tested two encodings:
Unary (as it is very close to the analytical results),
and Elias-$\gamma$.
The latter is better than unary because
for an integer $k$,
Elias-$\gamma$
requires
$2\lfloor \log_{2}k \rfloor +1$ bits,
while unary requires $k$ bits.


\begin{figure*}
\centering
    \includegraphics[width=2.4in]{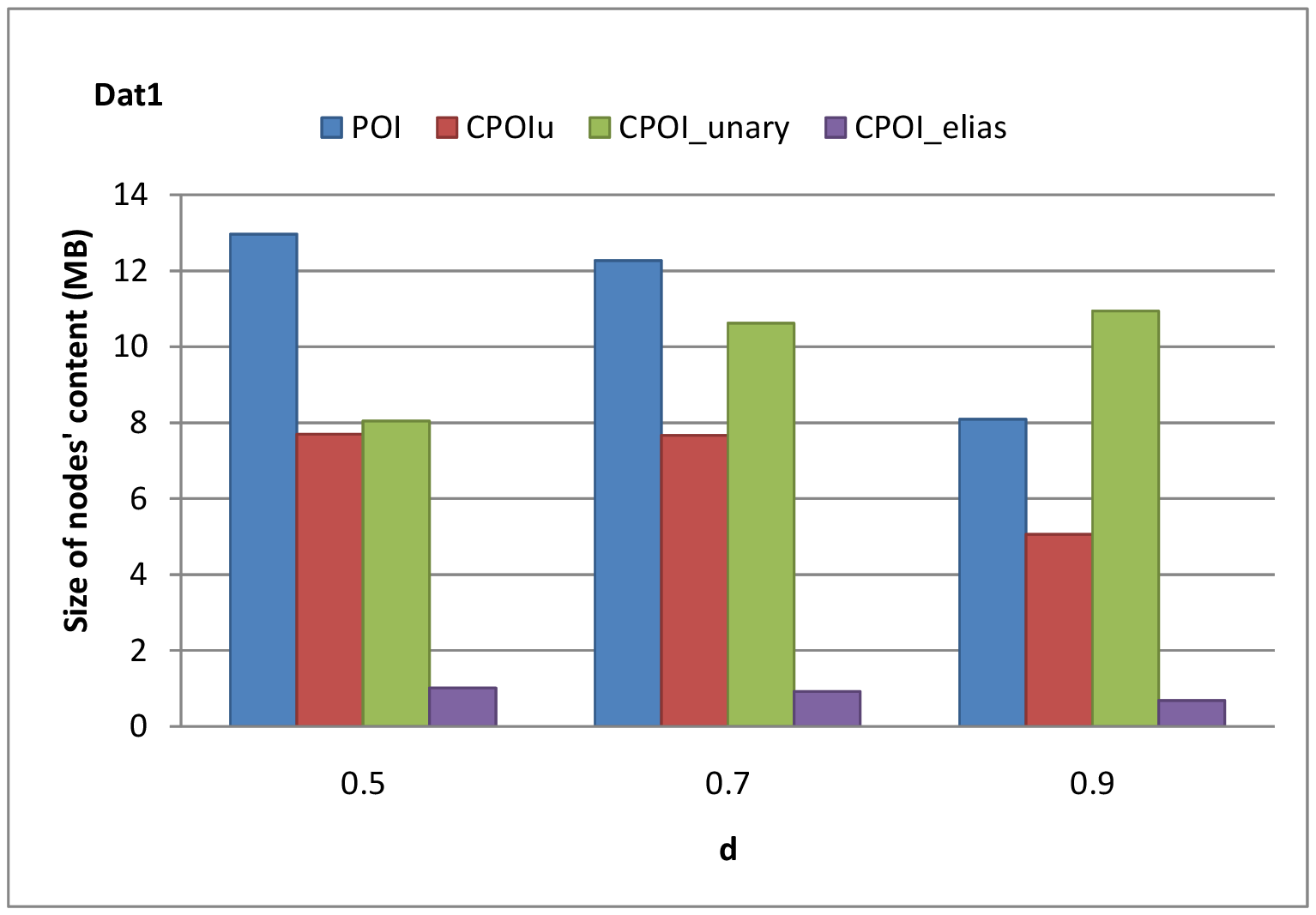}
    \hspace*{1cm}
    \includegraphics[width=2.4in]{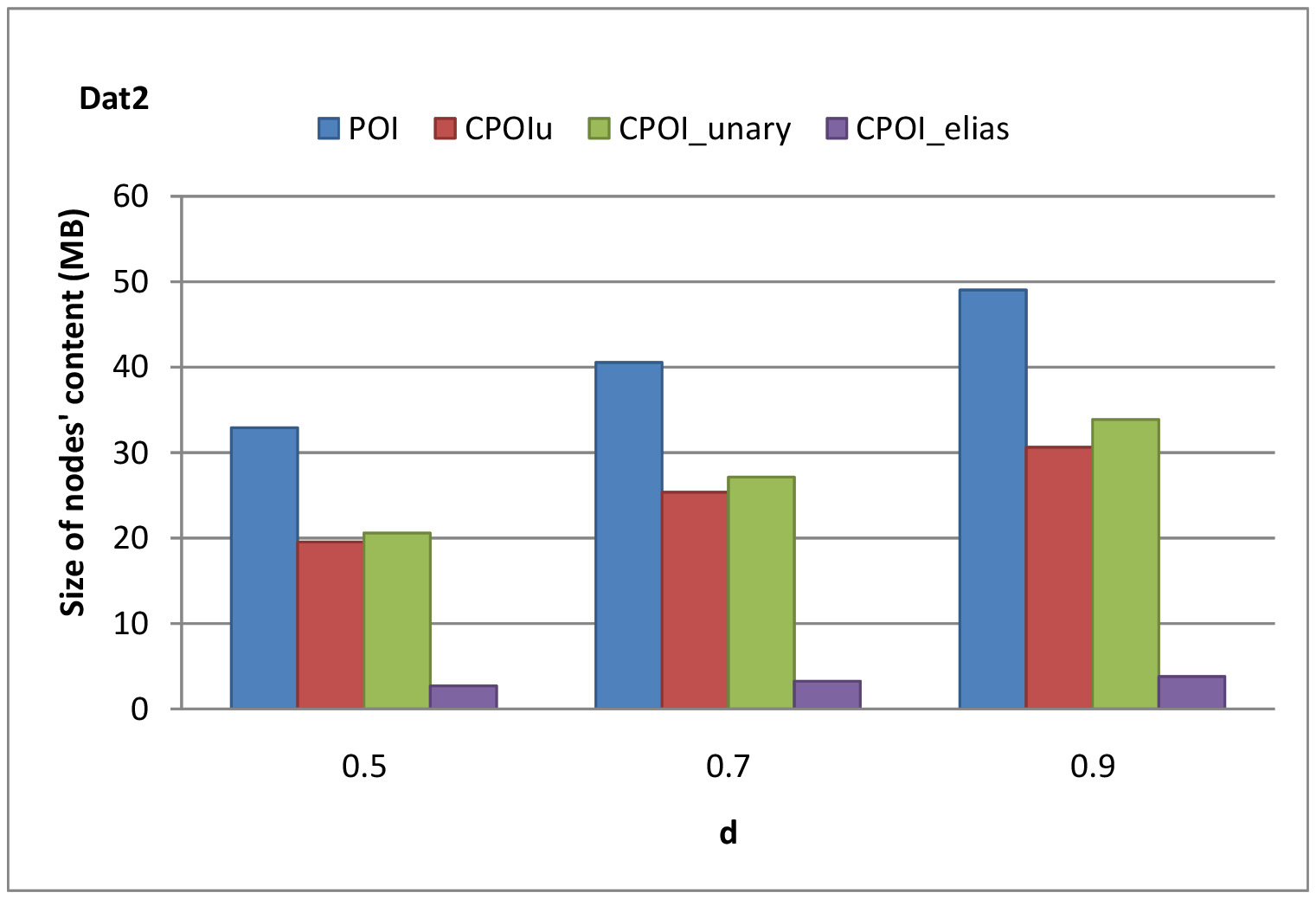}
\vspace*{-0.2cm}
\caption{Space for the nodes' contents for \DatA\ (left) and \DatB\ (right) of
        \poi, \cpoiu\ and   \cpoi\ (for Unary and Elias-$\gamma$ encoding).}
\vspace*{-0.3cm}
\label{fig:unaryAndElias}
\end{figure*}


\subsection{Experimental Results}

\noindent
[\DatA\ and \DatB]\\
The nodes' sizes  for \DatA\ (resp. \DatB) are shown
in the left (resp. right) part of Figure \ref{fig:unaryAndElias}.
It is evident that in both datasets
\cpoi\ with Elias-$\gamma$ is by far the best compression method,
achieving a 7.5\%-8.5\% compression ratio.
\cpoiu\ comes second achieving only 59\%-63\% compression ratio.
Third comes \cpoi\ with Unary encoding.

Regarding the $d$ value,
since in \DatB\ we have less subset or superset relations wrt \DatA,
as $d$ increases, $|T|$ and $\sum_{i=1}^{|N|}|n_{i}|$ increases as well,
consequently, the required space for \poi\ increases.
Regarding \cpoi, its space in \DatB\ increases,
as in \DatA,
since it depends on $Gaps(N)$ and for larger $d$ we have larger $|T|$.

\begin{figure*}
        \centering
            \includegraphics[width=3.12in]{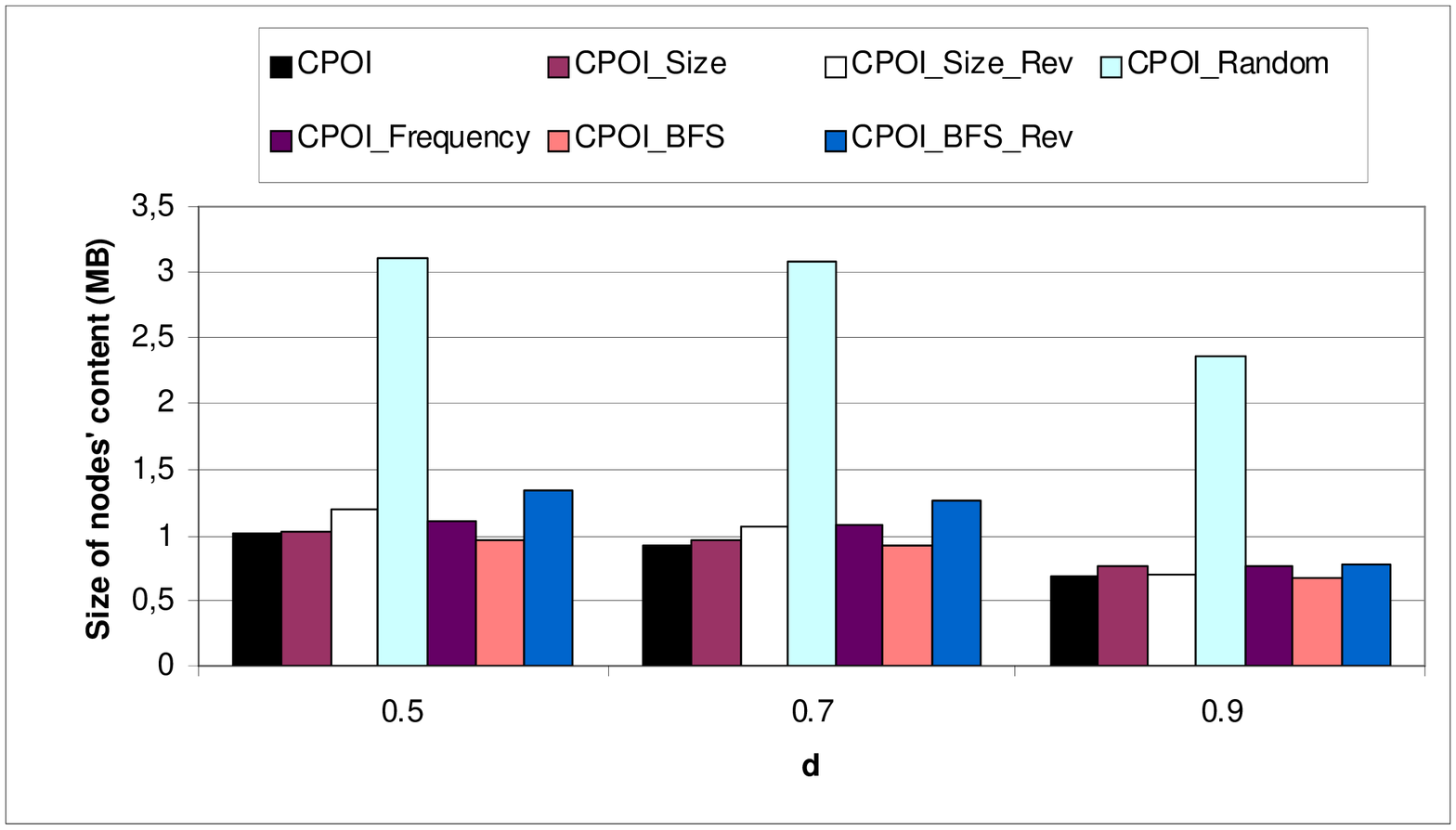}
            \includegraphics[width=3.12in]{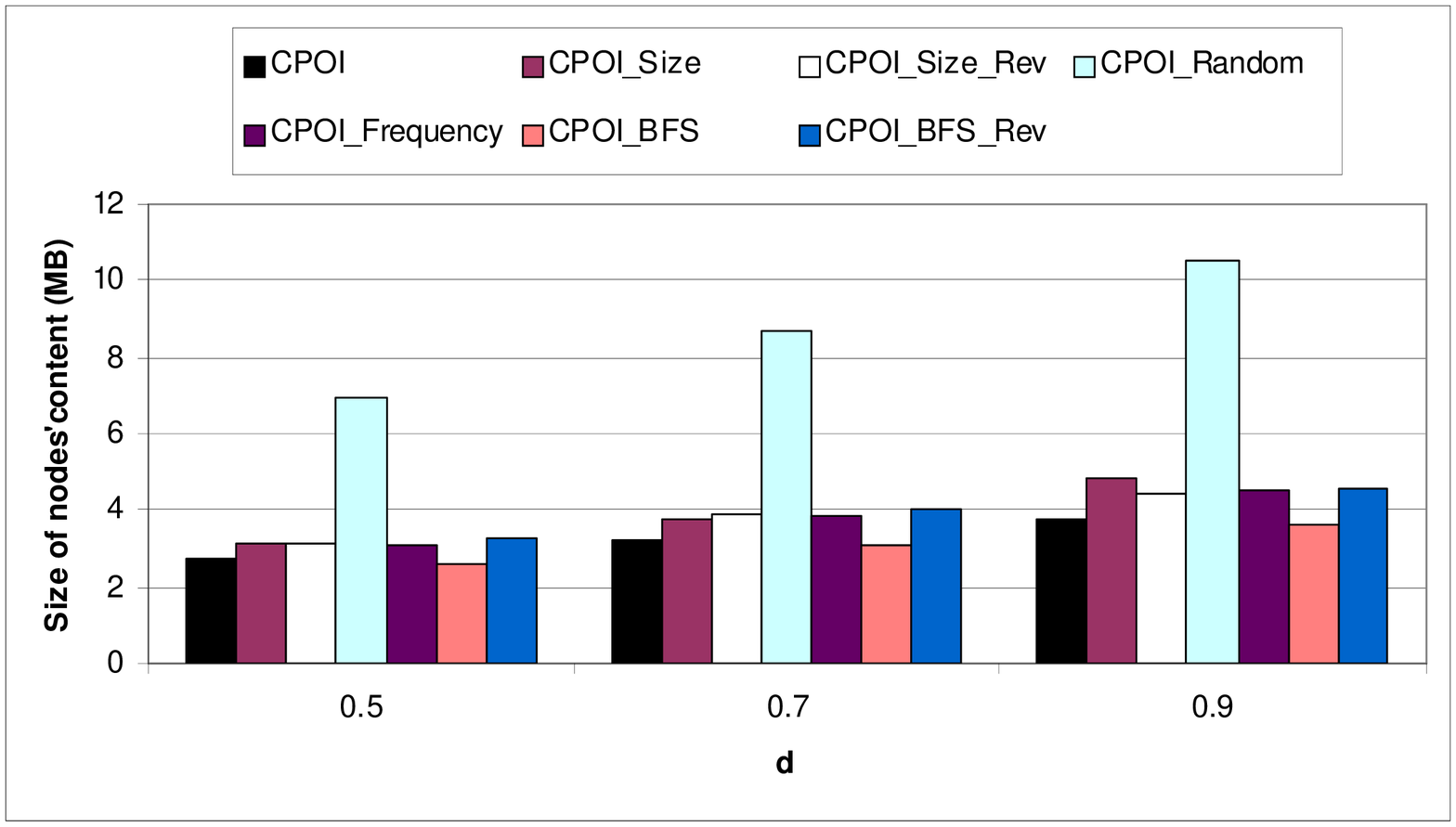}
        \vspace*{-0.2cm}
        \caption{Nodes's space for  \DatA\ (left) and \DatB\ (right) of \cpoi\
        using \underline{Elias-$\gamma$ encoding} for various assignment policies.}
        \vspace*{-0.2cm}
        \label{fig:spaceEliasGReassignment}
        \end{figure*}
\begin{figure}
\centering
    \includegraphics[width=2.9in]{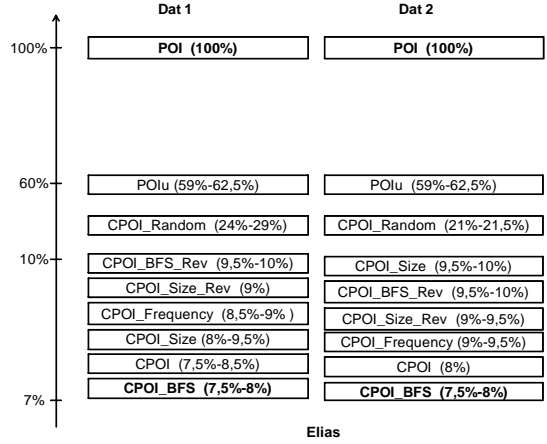}
\vspace*{-0.2cm}
\caption{Ranking of reassignment policies wrt their compression ratio for Elias-$\gamma$ encoding }
\vspace*{-0.3cm}
\label{fig:ReassignmentElias}
\end{figure}

Regarding the reassignment policies,
Figure \ref{fig:spaceEliasGReassignment} shows comparative
results for \DatA\ and \DatB\
using Elias-$\gamma$.
In both datasets \emph{BFS}
is the best reassignement
(with compression ratio 7.5\%-8\%),
slightly outperforming  \cpoi\ \emph{Default}.
The rest reassignment policies
are outperformed by \cpoi\, but none of them is worse than \poi\ or \cpoiu.
Figure \ref{fig:ReassignmentElias} groups  and ranks reassignment
policies according to the compression ratio they achieve for each dataset.

Another interesting observation
is that
a gapped representation with Elias-$\gamma$ gives around 20\%
compression ratio (even with a random id assignment).
A ``good" (i.e. at least not random) assignment
(with  Elias-$\gamma$ encoding) gives around 8\% compression ratio
(i.e. three times less space)
in comparison to the random assignment.

The time required for the reassignment
is short for all policies ranging from 4 to 11 secs for \DatA,
and from 14 to 46 secs for \DatB.\footnote{
    The implementation is in Java  and all experiments were carried
    out in a PC with Pentium(R) IV 3.40 Ghz, 1,49 GB Ram, and
    Windows XP.
}
The fastest policy is \emph{BFS}, while the slowest is \emph{Frequency}.


\begin{figure*}
\centering
    \includegraphics[width=2.7in]{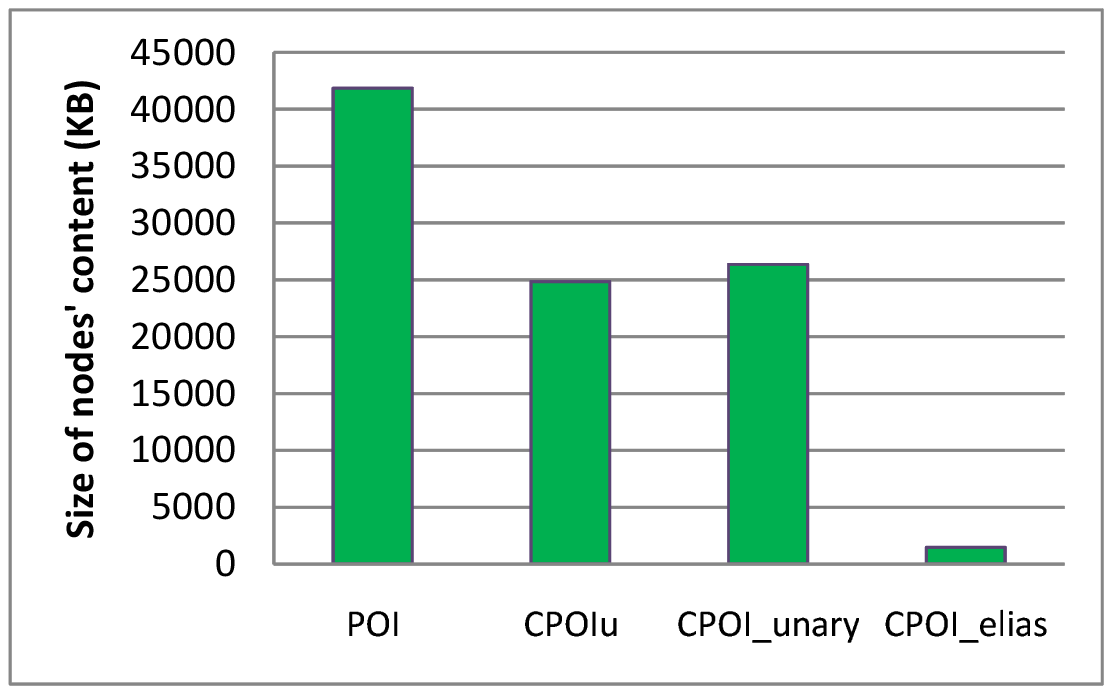}
    \includegraphics[width=2.8in]{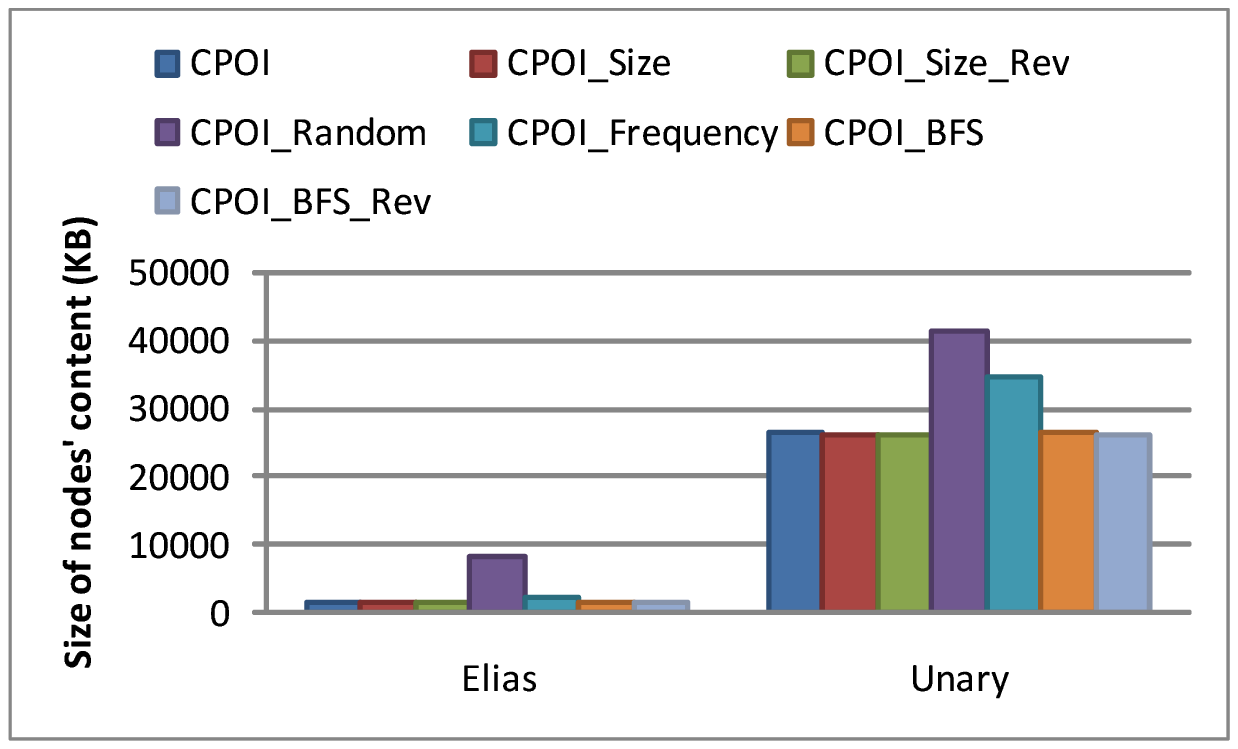}
\vspace*{-0.2cm}
\caption{Node space in \DatC.}
\vspace*{-0.3cm}
\label{fig:Dat3Space}
\end{figure*}

The main results can be summarized  as follows:
Using a \cpoi\ with \emph{BFS} reassignment
and Elias-$\gamma$
we can achieve compression ratio of 8\% of the size of
nodes' content.

\godown

\noindent
[\DatC] \\
The results for the nodes' space
are shown in the left
part of Figure \ref{fig:Dat3Space}.
We can see that
\cpoi\ outperforms \poi\ as we expected.
\cpoi\ with Elias-$\gamma$ gives
the best compression ratio, i.e.  3.5\%.
\cpoiu\ comes second achieving 59.4\% compression ratio,
while \cpoi\ with Unary encoding achieves 63\% compression ratio.
Regarding the reassignment policies,
Figure \ref{fig:Dat3Space} (right)
shows comparative
results using Elias-$\gamma$ and unary encoding.
The best reassignement is \emph{Size\_Rev} for unary and \emph{Size} for Elias-$\gamma$
(with compression ratio 62\% and 3.3\% respectively).
\emph{Size}, \emph{Size\_Rev} and \emph{BFS\_Rev}
outperform  \cpoi\ \emph{Default}.
The other reassignment policies
are outperformed by \cpoi\ \emph{Default}, but none of them is worse than \poi\ or \cpoiu.

\godown
\noindent
[GO Dataset]\\
%
%
In this dataset, although the graph of \poi\ is again flat,
\cpoi\ is better than \cb\ and \poi\
(detailed results are given in the Appendix \ref{sec:TotalSpace}).


\ \\
\subsection{Experimental Results vs Analytical Results}

Here we discuss the datasets and the experimental results
under the light of the analysis of Section \ref{sec:Approach}.
The conditions that do not require computing $Gaps(N)$,
i.e. those
in
Prop. \ref{prop:Avg} and equations
(\ref{eq:cpoiAlwaysBetterThanCpoiu}) and (\ref{eq:cpoiAlwaysWorstThanCpoiu}),
are not satisfied,
in none of \{\DatA, \DatB, \DatC\},
hence
by looking at the features of our datasets
(i.e. $|n_i|, |T|, avg |n_i|$)
we cannot conclude
which approach guarantees space benefits.
However, in the GO dataset, the conditions of
Prop. \ref{prop:Avg} and equation
    (\ref{eq:cpoiAlwaysBetterThanCpoiu}) hold,
    so \cpoi\ always
    guarantee space savings over both \poi\ and \cpoiu.
    Consequently, the conditions of Prop. \ref{prop:Gain} and \ref{prop:Uniform}, hold too for GO.

Referring to the conditions that require a gap representation
(i.e. those in Prop. \ref{prop:Gain} and Prop. \ref{prop:Uniform}),
by considering the \emph{Default} reassignment,
Prop. \ref{prop:Gain} holds for \DatA\ for all $d\in [0.5-0.7]$,
while for \DatB\ for $d\in [0.5-0.9]$.
Prop. \ref{prop:Uniform}
does not hold for any value of $d$ neither for \DatA\ nor for \DatB.
Regarding, the other reassignment policies
Prop. \ref{prop:Gain} holds for \DatB\ for \emph{BFS}, \emph{Frequency} and \emph{Size\_Rev},
while
for \emph{Size} and \emph{BFS\_Rev} it holds only for $d=0.5$.
In \DatC\ only Prop. \ref{prop:Gain} holds for all reassignment policies.

These results agree  with the experimental results
(i.e. the satisfied inequalities also hold
in the measured experimental results).

\subsubsection*{CBD (CB Dictionary-Based approach)}

\begin{figure*}
\centering
    \includegraphics[width=115mm]{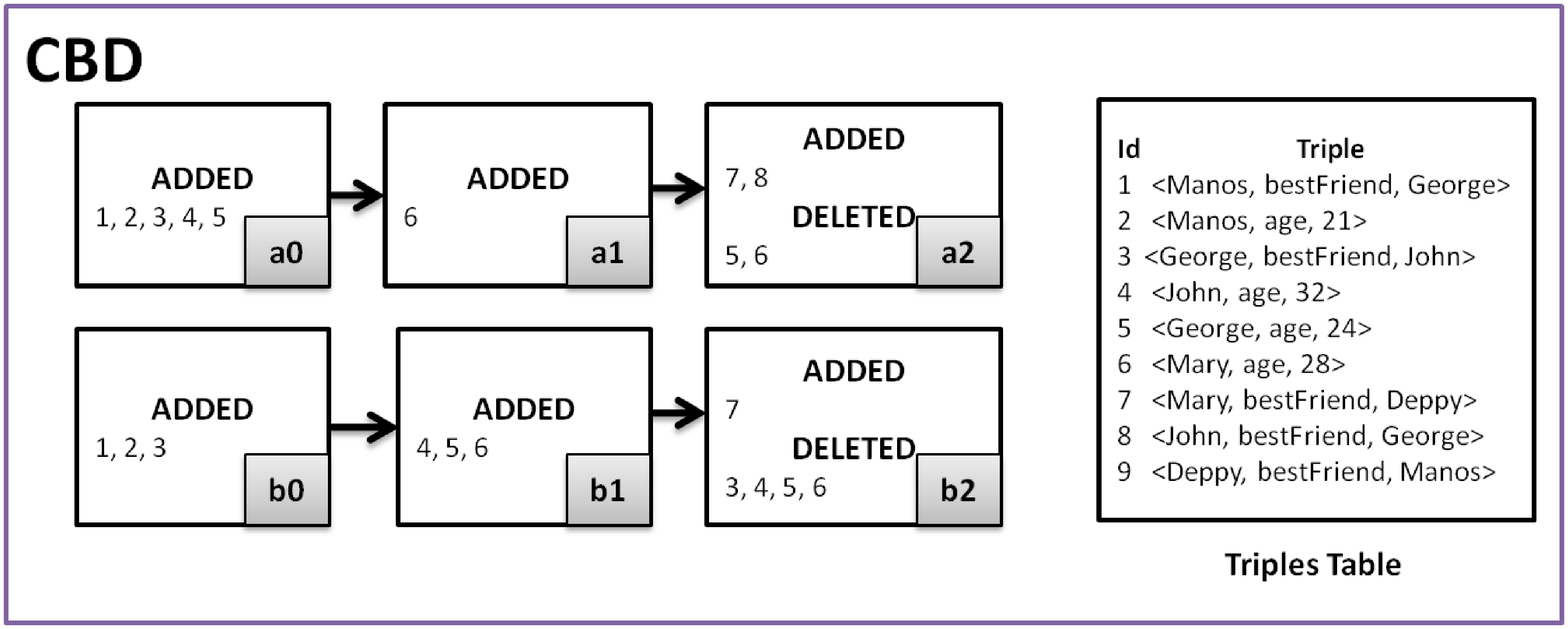}
\vspace*{-0.2cm}
\hspace*{0.9cm}
\caption{Storing versions according to the \cbd\ approach}
\label{fig:CBD_Structure}
\vspace*{-0.6cm}
\end{figure*}

We have
included in our experiments
also
 a {\em change-based dictionary approach}, for short \cbd,
in which
all deltas contain additions and deletions of {\em identifiers} rather than
triples.
Figure \ref{fig:CBD_Structure} depicts what
the \cbd\
approach would store for the versions of the example
of Fig. \ref{fig:tripleTable}-\ref{fig:POI_Structure}.
In general, this
approach is certainly better than \cb\ for those
triples of $T$ that occur at least in two deltas.
In GO \cbd\ is worse than \cb.
In \DatC\
\cbd\ behaves much better than \cb,
but worse than \cpoi,
as shown
in the Appendix \ref{sec:TotalSpace}.
The same happens in \DatB.


\subsection{General Discussion \& Version  Insertion Times}


Regarding the comparison
of \poi\ (and \cpoi) with \cbd,
in \cbd\
the cost to reconstruct the contents of a version is history dependent
(and may more expensive than \poi),
and \poi\ offers faster subset checking.
On the other hand, the addition
of a version in \cbd\ can be faster than \poi\
since there is no need for checks to
determine its placement in the storage graph.
%
In general we can say that the price to pay,
in comparison  to \ic\ and \cb\ (as well as \cbd),
is slower additions of new versions
and the adoption of gapped and encoded identifiers
makes them slower.
Although the insertion algorithm of \poi\ exploits the structure and semantics of the storage graph,
its plain version sometimes
requires
a long time
for the addition of a version.

However, the variation with cache
which has been proposed in \cite{DBLP:conf/esws/TzitzikasTA08},
reuses results of set union operations,
and leads to an average insertion time {\em less than one second}.

In our experiments, the average insertion time for \DatA\ (resp. \DatB) was 0.3-0.8 sec (resp. 0.8-1 sec).
The gapped and encoded identifiers indeed incurred  an overhead,
specifically in our experiments for \DatA\ (resp. \DatB) the insertion time was
0.8-2.4 sec (resp. 3.2-4.4 sec)
which is however an {\em acceptable time}.
To be more specific, consider a version with $n$ triples.
For inserting that version to \cpoi,
the extra cost  (in comparison to \poi)
is the cost
to decode the gapped identifiers
of those nodes
that will be examined
for deciding the right place of
the new node that it may have to be inserted in the lattice of \cpoi.
The extra cost for decoding the identifiers of a node with $n$ triples
is linear (in $\OU(n)$)
since we have to perform at most one integer addition
operation per triple.

Finally we should stress
that since a version is added only {\em once},
but it can be retrieved {\em unlimited} number of times,
\cpoi\ is a beneficial choice
for the intended application scenario.

\section{Possible Applications}
\label{sec:Appls}

Here we describe in brief possible application contexts of \cpoi.

\bi
\item {\em Generic versioning system for RDF/S datasets} \\
  Development for versioning systems (server and client) for RDF/S triple sets,
  like SVN.
  The server's side storage can be  based on \cpoi.

\item {\em Domain-specific  archiving solutions}. \\
  In various domains, e.g. in the Marine domain,
  primary data are kept and updated in relational databases
  and are periodically exported in RDF/S format
  according to the LOD (Linked Open Data) best practices.\footnote{
        This is approach adopted
        in ECOSPOPE  (http://www.ecoscopebc.ird.fr/),
        or in various
        departments
        of
        Food and Agricultrure Organization of the United Nations
        (FAO UN),
        e.g.
        the FLOD KB
        of the
        Fishery and Aquaculture department
        (http://www.fao.org/fishery/topic/18046/en).
  }
  As consequence of this process,
  past versions of the RDF/S datasets are lost.
  \cpoi\ can  be exploited for versioning
  these RDF/S datasets
  mainly those  comprised of scientific data over which
  other experiments take place.
  In this way
  the operational database (which is subject of updates)
  does not need to be enriched with any versioning services.

  In general, one general approach is to adopt a system composed of
  three subsystems:
  (a) the operational (transactional) database where updates take place,
  (b) the latest dataset exported in RDF/S and indexed
      for offering fast (SPARQL) query services
     (e.g. it may be stored in a system like RDF-3X \cite{DBLP:journals/pvldb/NeumannW08}),
  (c) a system based on \cpoi\ that archives and makes accessible the past versions.
  Note that in case one agent (human or application) would like to browse
  or query a past version
  that version can be retrieved from \cpoi\
  and expressed in a browsable or queryable system.

    Although the focus on \cpoi\ is archiving, and not
time-travel queries or cross version operations,
we should note that some operations are very easy
and fast to perform over \cpoi.
Specifically, it is fairly easy and fast to
find all versions whose contents are subset (or superset) of a given version,
all versions that include or are subset of a given set of triples, etc.
\ei


\section{Conclusion}\label{sec:Conclusions}


We proposed a compact representation for \poi\,
based on gapped representation of triple identifiers
and variable-length identifier encodings.
We analyzed the space requirements of this representation
and identified sufficient conditions that guarantee compression
comparing to plain \poi.
%
Subsequently, we conducted a large  number of experiments
over various synthetic datasets, and using several methods for assigning identifiers.
The experimental results can be  summarized  as follows.
Regarding identifier assignment policies,
we noticed that an  assignment
in a {\em first-in-first-served} basis
is almost as good
as
reassignment policies,
and from this
we can conclude that identifier reassignment
is not necessary.
Using a \cpoi\
we can achieve a compression ratio
(in comparison to plain \poi)
of about 8\% of the size of nodes' content.
Note that the adoption of a uniform representation (i.e. \cpoiu)
would achieve the same compression ratio of 8\% if:
$
\frac{Space_{CPOI_U}(N)}{Space_{POI}(N)}=8\% \Leftrightarrow
\frac{\lceil \log_{2}|T|\rceil}{\B}=8\%
\stackrel{\B=32}{\Leftrightarrow}
\lceil \log_{2}|T| \rceil = 2.56
\Leftrightarrow   |T|=6$,
i.e.
if the number of distinct identifiers
were only $6$,
%
however in our experiments the number of distinct triples were more than half a million!
This demonstrates the benefits of the gapped and specially encoded identifiers.

We have also seen
that even in anti-correlated for \poi\  datasets,
\cpoi\ is still better than the \cb\ approach.
Finally, we should
stress that since
we do not deal with the compression of the table that keeps
the distinct triple strings and their ids,
techniques like those proposed
in \cite{DBLP:journals/pvldb/NeumannW08} and \cite{iswc/Fernandez10},
are complementary to \cpoi\
and if they are used {\em together} will further reduce the overall  space.

The price to pay is slower (than \ic\ or \cb) insertion times.
However note that although
a version can be  retrieved unlimited number of times,
it can be added (i.e. inserted to \cpoi) only {\em once}.
Therefore \cpoi\ is a beneficial choice
for the intended application scenario.

At last we should note
that apart from RDF/S datasets,
\cpoi\
can be a beneficial choice
for archiving sets
of identifiers.
For instance,
    social networking systems have to keep information about
    large numbers of users and their memberships to various groups.
    \cpoi\ could be exploited for versioning such groups.

\small

\bibliographystyle{plain}
\bibliography{diffBibliography,paper}


\normalsize
\appendix
\section{Proofs}

The proofs of some propositions follow
directly from the discussion
and therefore are omitted.

\noindent
{\bf Prop. \ref{prop:AllDisjoint}}.
If a storage graph has $|N|$ sets
and they are \emph{pairwise disjoint}, then:
$
 |T|-|N| \leq Gaps(N) \leq |N|(|T|-|N|)
$.

\noindent
Proof: \\
If all sets of $N$ are {\em pairwise disjoint}
(i.e. $n_{i} \cap n_{j} = \emptyset$,  $\forall i\neq j,$ $i,j\in [1,|n|]$),
then we can achieve the best assignment of identifiers  for every node
by assigning consecutive ids to the triples of every distinct node. Hence:

{\small
\begin{equation}
 Gaps(N) \geq \sum_{n_{i}\in N}(|n_{i}|-1) = \sum_{n_{i}\in N}|n_{i}|-\sum_{n_{i}\in N}1 = |T|-|N|
\end{equation}
}

On the other hand,
the worst assignment
in the above case
is the one obtained when every node contains triples (ids) that cover the greatest possible range of values.
Specifically, the first node will contain the triples with ids 1 and $|T|$
(these will be the min and max ids of that node),
and hence, $gaps(n_{1})=|T|-1$ (according to the worst case of a node).
Respectively, the second node will include the triples with ids 2 and $|T|-1$
(since every triple occurs only once), and hence, $gaps(n_{2})=(|T|-1)-2=|T|-3$.
We can proceed analogously for the rest nodes,
so the $k$-th node will contain the triples with ids $k$ and $|T|-(k-1)$,
and thus  $gaps(n_{k})=(|T|-(k-1))-k=|T|-(2k-1)$.
Therefore,
$Gaps(N) \leq (|T|-1)+(|T|-3)+...+(|T|-(2|N|-1))$.
%
%
{\small
\begin{eqnarray*}
 Gaps(N)&\leq&(|T|-1)+(|T|-3)+...+(|T|-(2|N|-1))  \\
            &=&  \sum_{i=1}^{|N|}(|T|-(2i-1))
            =  \sum_{i=1}^{|N|}(|T|) - \sum_{i=1}^{|N|}(2i-1) \\
            &=&  |N||T| - 2\sum_{i=1}^{|N|}i + |N| \\
            &=&   |N||T| - 2\frac{|N|(|N|+1)}{2} + |N| \Leftrightarrow \\
 Gaps(N) &\leq & |N|(|T|-|N|)
\end{eqnarray*}
}
$\diamond$


\noindent
{\bf Prop. \ref{prop:Overlaps}}.
If a storage graph has $|N|$ sets
and there are \emph{overlaps} then:
$
\sum_{i=1}^{|N|}|n_{i}|-|N| \leq Gaps(N) \leq |N||T|-|N|
$.

\noindent
Proof:\\
Let's consider the case that leads to the worst reassignment for
every node.
The worst reassignment  occurs when every
node  contains triples (ids) that cover the whole range of values.
Specifically, for each node $n_{i}$ we have $gaps(n_{i}) = |T|-1$.
Note that this case can occur only if the intersection of all nodes
$n\in N$ is greater or equal than 2. Consequently,
$ Gaps(N) \leq |N|(|T|-1) =  |N||T|-|N|$.
On the other hand, the case that could lead to the best reassignment
if overlaps exist, occurs when the intersection between two nodes
consists of triples with consecutive ids.
Indeed in the first node those ids should be at the beginning of the list,
while in the second
they should be at the end. For instance, consider four nodes:
$n_{1} = \{1, 2, 3\}$, $n_{2} = \{2, 3, 4\}$,
$n_{3} = \{3, 4, 5, 6\}$, $n_{4} = \{5, 6, 7\}$. In
that case, we can achieve consecutive ids that differ by one, hence:
\begin{eqnarray*}
 Gaps(N) \geq \sum_{i=1}^{|N|}(|n_{i}|-1) = \sum_{i=1}^{|N|}|n_{i}|-|N|
\end{eqnarray*}
We conclude that in the case that there are overlaps among the nodes' contents then
$
\sum_{i=1}^{|N|}|n_{i}|-|N| \leq Gaps(N) \leq |N||T|-|N|
$.\\
$\diamond$

\noindent
{\bf Prop. \ref{prop:Avg}}.
 If  the number of distinct triples
 is not greater than $\B$ times the average number of elements of a node,
 then \cpoi\ saves space.

\noindent
Proof: \\
By combining Prop. \ref{prop:Overlaps} and Prop. \ref{prop:Gain}
it follows  that we gain with a \cpoi\ if
the upper bound of $Gaps(N)$ according to Prop. \ref{prop:Overlaps}
is less
than the right hand of the condition of Prop. \ref{prop:Gain} (which guarantees gain),
i.e. if :
right of Prop. \ref{prop:Gain} $>$  right of Prop. \ref{prop:Overlaps}.

right of Prop. \ref{prop:Gain} $>$  right of Prop. \ref{prop:Overlaps} $\Leftrightarrow$
\begin{eqnarray*}
 && \B*\sum_{i=1}^{|N|}(|n_{i}|-1) >  |N||T|-|N| \Leftrightarrow \\
 && \B*\sum_{i=1}^{|N|}|n_{i}| - \B*|N| - |N||T| + |N| > 0  \Leftrightarrow \\
 && \B*\sum_{i=1}^{|N|}|n_{i}| - |N|*(|T| + \B-1) > 0
\end{eqnarray*}

By expressing $\sum_{i=1}^{|N|}|n_{i}|$ as $|N|*avg(|n_i|)$,
where $avg(|n_i|)$ is the average number of elements of a node,
we can write:
$
 \B*|N|*avg(|n_i|) - |N|*(|T| + \B-1) > 0 \Leftrightarrow
 \B*avg(|n_i|) > |T| + \B-1
$.
The above says that we can always save space with
a \cpoi\ when the number of distinct triples is not greater than $\B$ times
the average number of elements of a node.

\ \\
$\diamond$

\noindent
{\bf Prop.} \ref{prop:Uniform}. {\em
\cpoi\ requires less space than \cpoi$_U$,
iff
$Gaps(N) \leq \sum_{i=1}^{|N|}|n_{i}| * \lceil \log_{2}|T| \rceil - \B*|N|$.
$\diamond$
}

\noindent
Proof: \\
\cpoi\ requires less space
than \cpoiu\ when:
\begin{eqnarray*}
 \B*|N| + Gaps(N) \leq \sum_{i=1}^{|N|}|n_{i}| * \lceil \log_{2}|T| \rceil \Leftrightarrow \\
 Gaps(N) \leq \sum_{i=1}^{|N|}|n_{i}| * \lceil \log_{2}|T| \rceil - \B*|N|
\end{eqnarray*}

\ \\
$\diamond$


\noindent
{\bf Prop. \ref{prop:WUBUE}}.
The worst case of unary is better than uniform encoding, when:
$avg(|n_i|) * \lceil \log_{2}|T| \rceil \geq |T|+\B-1$

\noindent
Proof:
\begin{eqnarray*}
 \B*|N| + |N|*|T| - |N| \leq  \sum_{i=1}^{|N|}|n_{i}| * \lceil \log_{2}|T| \rceil \Leftrightarrow \\
 \sum_{i=1}^{|N|}|n_{i}| * \lceil \log_{2}|T| \rceil \geq |N|*(|T|+\B-1)  \Leftrightarrow \\
avg(|n_i|) * \lceil \log_{2}|T| \rceil \geq |T|+\B-1
\end{eqnarray*}
\ \\
$\diamond$


{\bf Prop. \ref{prop:BUWUE}.}
The best case of unary is worse than uniform encoding,
and therefore
uniform is certainly better than unary representation,
when:
$
avg(|n_i|) * (\lceil \log_{2}|T| \rceil -1) \leq \B-1
$

\noindent
Proof:
\begin{eqnarray*}
 \B*|N| + \sum_{i=1}^{|N|}|n_{i}| - |N| \geq  \sum_{i=1}^{|N|}|n_{i}| * \lceil \log_{2}|T| \rceil \Leftrightarrow \\
 (\B-1)*|N| \geq \sum_{i=1}^{|N|}|n_{i}| * (\lceil \log_{2}|T| \rceil -1) \Leftrightarrow \\
 avg(|n_i|) * (\lceil \log_{2}|T| \rceil -1) \leq \B-1
\end{eqnarray*}
$\diamond$

\section{Total Space}
\label{sec:TotalSpace}

The benefits of \cpoi\ are independent on the
size of {\em triple strings},
in the sense that the nodes of \poi\ and \cpoi\
store triple {\em indentifiers}.
However,
here we report the results
of measurements
that include the size of the triple strings
just for giving the reader the complete picture.
However,
and as we stressed in the main body of the paper,
one could adopt other complementary techniques
for compressing the triple strings themselves
(i.e. assign one id for each subject, triple, object of a triple).
This means that one could achieve even better compression ratios
than those that we report in this section.

\godown
\noindent
[\DatA\ and \DatB] \\
        To quantify the overall benefit of using \cpoi,
        we compared the total storage space requirements of \ic, \cb, \poi,
        \cpoi\ (Elias-$\gamma$ and \emph{Default} id assignment),
        and \cpoi\ with the best reassignment policy wrt the experiments
        (i.e. \emph{BFS}).
        For \cb\ we used the symmetric difference operator
        ($\Delta_e$ in \cite{Zeginis2007}),
        i.e.
        the difference between two sets of triples $A$ and $B$,
        is the set
        $(A \setminus B) \cup (B \setminus A)$.

\begin{figure}
        \centering
            \includegraphics[width=75mm]{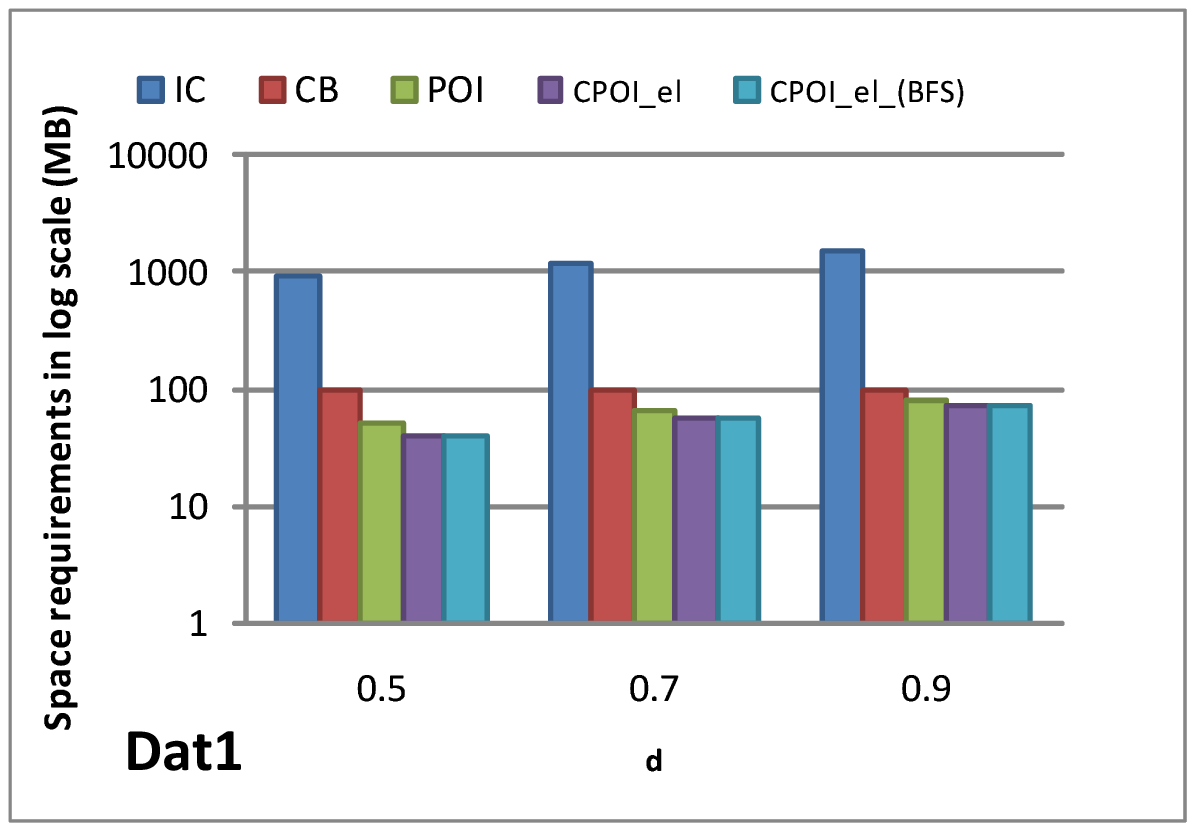}
            \includegraphics[width=75mm]{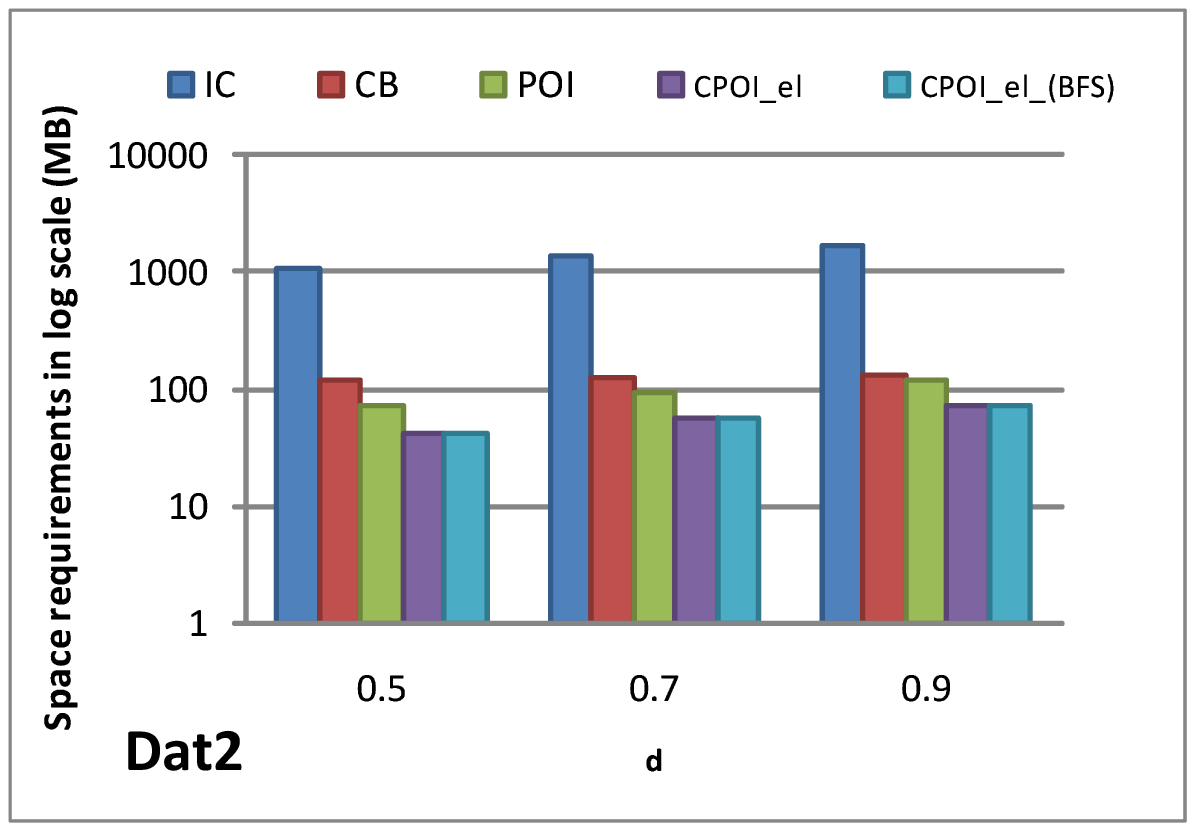}
            \includegraphics[width=75mm]{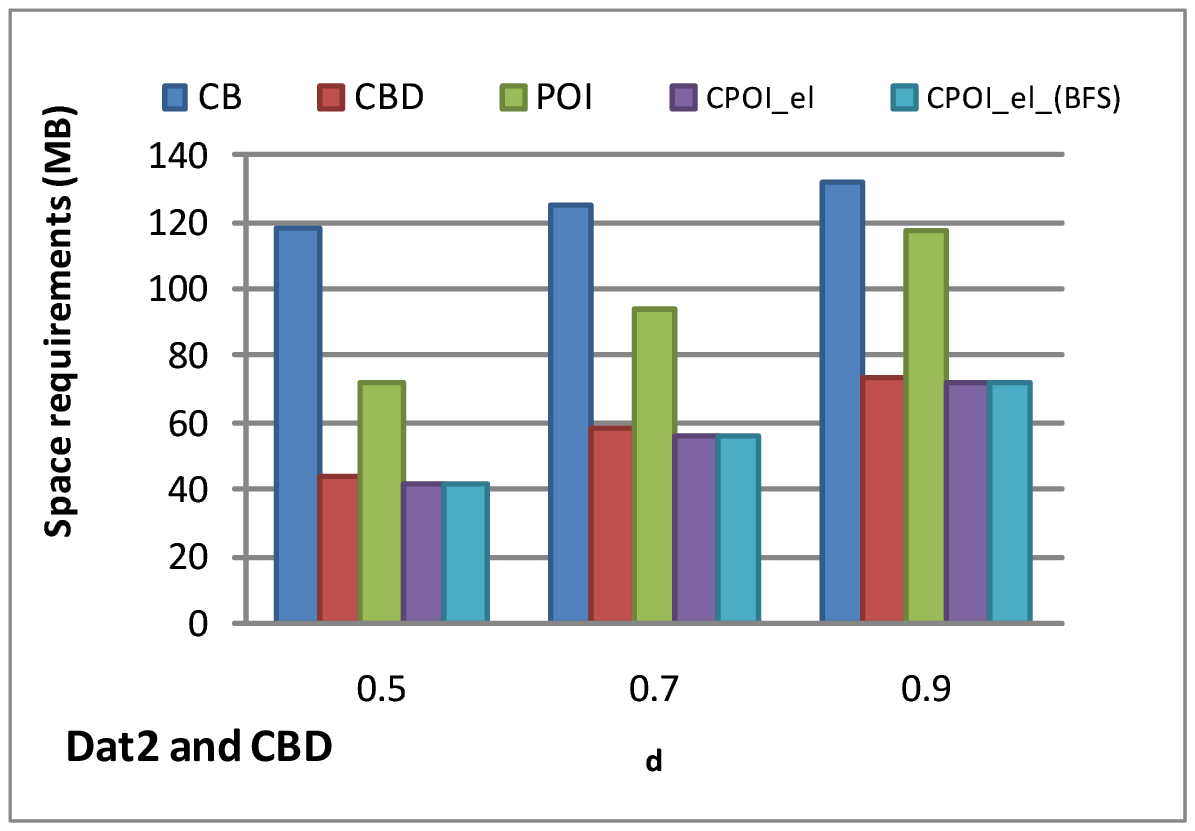}
        \vspace*{-0.2cm}
        \caption{Total space  for \DatA (up) and  \DatB\ (middle) of \ic, \cb, \poi,
        for various $d$ values. Bottom: \DatB\ and \cbd}
        \label{fig:totalBytes}
        \end{figure}

        The upper left (resp. right) part of Figure \ref{fig:totalBytes}
        shows the space requirements
        for \DatA\ (resp. \DatB) in \underline{log scale},
        for various values of $d$ (0.5 - 0.9).
        We can see that \cpoi\ is always better
        (the plots of \cpoi\ \emph{Default} and \cpoi\ \emph{BFS}
            coincide since the log scale reduces
            their difference).

Comparing the total size of \cpoi\
    with the other two methods (\ic\ and \cb),
    the former requires only
    about 4.3\% (4.4\%-4.8\% for \DatA\ and 3.8\%-4.3\% for \DatB) of the
    space needed
    for \ic\,
    and  35.4\%-70.9\% (40.1\%-70.9\% for \DatA\ and
    35.4\%-54.8\% for \DatB) of the
    space needed for \cb\ approach.

\begin{figure}
\centering
    \includegraphics[width=63mm]{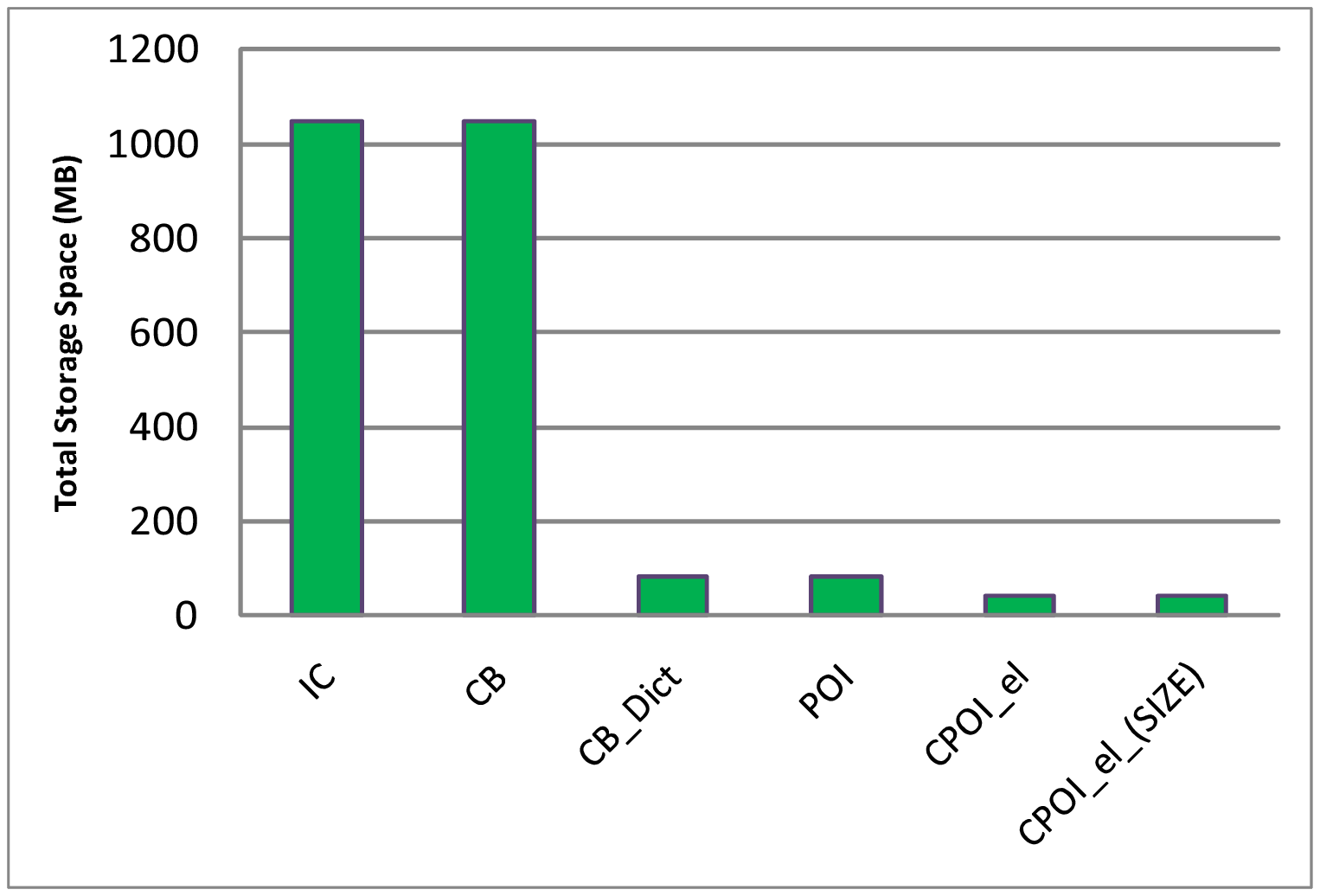}
\vspace*{-0.2cm}
\caption{Total storage space in \DatC.}
\vspace*{-0.3cm}
\label{fig:Dat3SpaceTotal}
\end{figure}

\godown
\noindent
[\DatC] \\
For this dataset the results regarding total  space
are shown in the Fig. \ref{fig:Dat3SpaceTotal}.
We can see that \poi\ is much better than both \ic\ and \cb\
(as it stores each distinct triple once),
while \cb\ is slightly worse than \ic.

We conclude that even if no version is subset of another,
and we have a significant number of versions,
then \cpoi\ is {\em significantly better} than the \cb\ approach.

\begin{figure}
\centering
    \includegraphics[width=63mm]{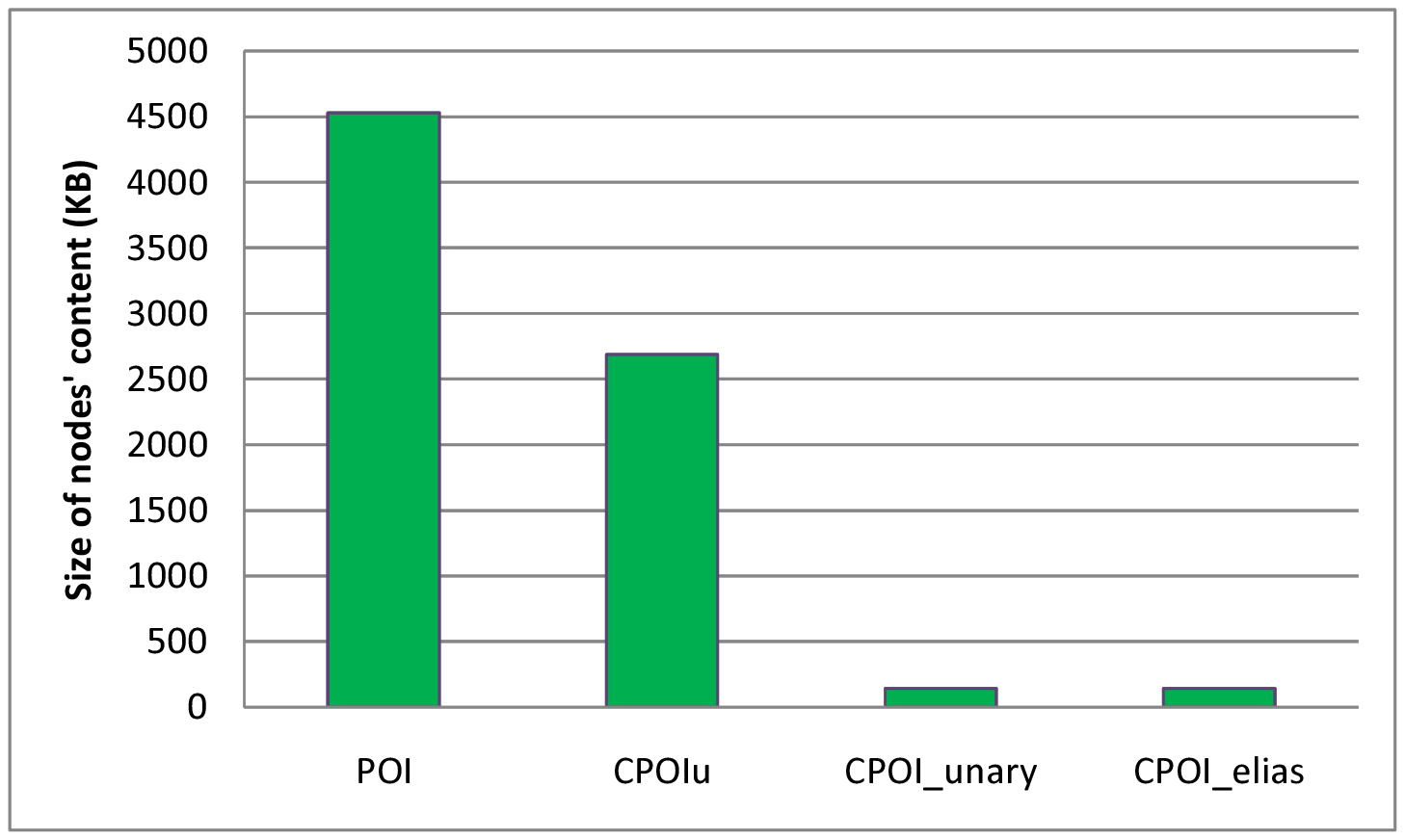}
    \includegraphics[width=63mm]{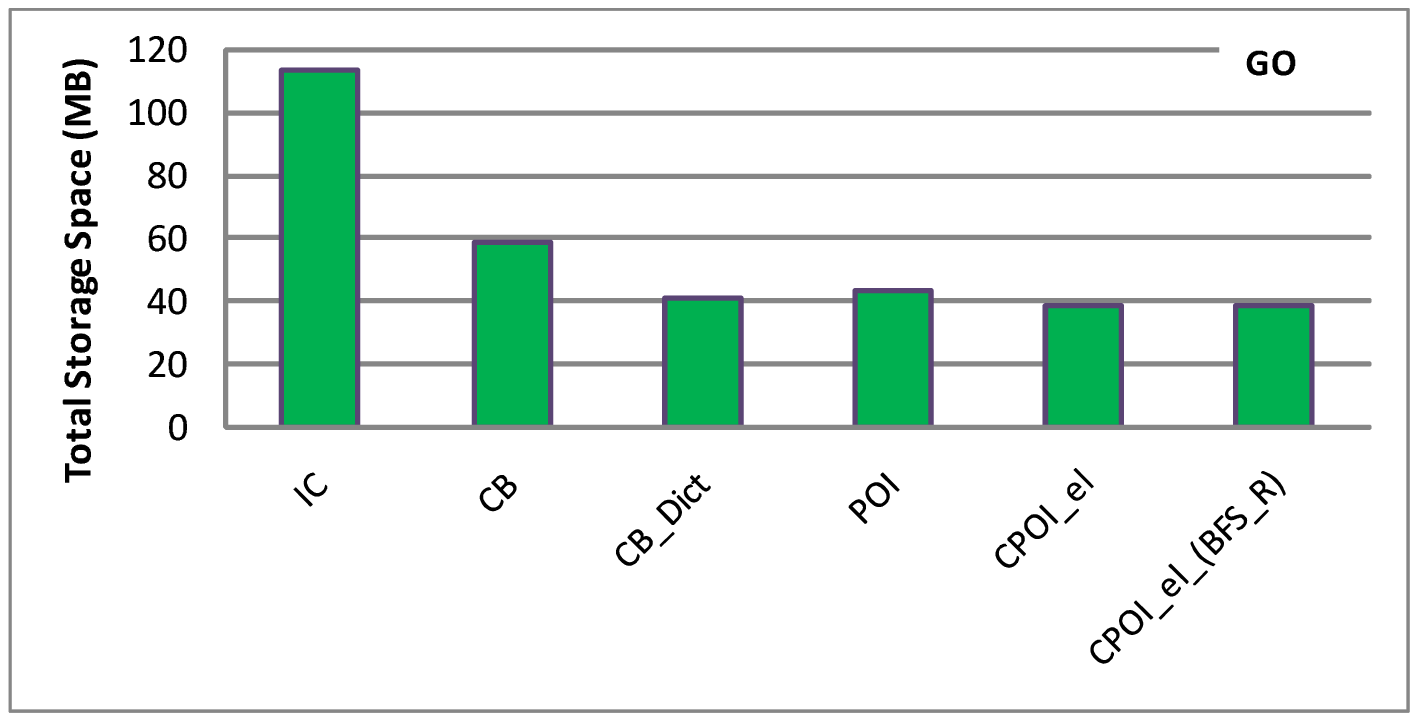}
\vspace*{-0.2cm}
\caption{Nodes space and total space in GO}
\vspace*{-0.2cm}
\label{fig:GO}
\end{figure}

\godown
\noindent
[GO Dataset]\\
Figure \ref{fig:GO} (up) shows comparative results
regarding the  storage space of nodes.
We observe that unary and Elias-$\gamma$ encodings
offer significant gains.
Now Figure \ref{fig:GO} (down) shows
the total storage space.
We can conclude that even
    with a few and not subset-related versions,
    \cpoi\ can be as good as the \cb\ approach
    (something which is very interesting).
Recall that this dataset is ideal
for
testing
the  worst case for \cpoi\
since no version is subset or superset of any other version
(therefore  the graph of \cpoi\ is flat),
and the number of versions is very small.

\end{document}